\documentclass[]{nature}
\makeatletter\if@twocolumn\PassOptionsToPackage{switch}{lineno}\else\fi\makeatother

\usepackage{tabulary,graphicx,amsmath,amsfonts,amssymb}
\usepackage[utf8x]{inputenc}
\usepackage[T1]{fontenc}
\usepackage{authblk}
\usepackage{abstract}
\usepackage{subfigure}
\usepackage{color}
\usepackage{hyperref}
\usepackage[table]{xcolor}
\usepackage{tabularx}
\usepackage{hhline}
\usepackage{epstopdf}
\usepackage{amsmath} 
\usepackage{float}
\usepackage{aliascnt}
\usepackage{amssymb}
\usepackage{upgreek}
\usepackage{lineno}

\makeatletter
\renewenvironment{figure}
               {\@float{figure}}
               {\end@float}
\renewenvironment{figure*}
               {\@dblfloat{figure}}
               {\end@dblfloat}

\renewenvironment{table*}
               {\@dblfloat{table}}
               {\end@dblfloat}

\makeatother

\usepackage{url,multirow,morefloats,floatflt,cancel,tfrupee}
\makeatletter

\AtBeginDocument{\@ifpackageloaded{textcomp}{}{\usepackage{textcomp}}}
\makeatother
\usepackage{colortbl}
\usepackage{xcolor}
\usepackage{pifont}
\usepackage{cite}
\usepackage[nointegrals]{wasysym}
\makeatletter

\def\mcWidth#1{\csname TY@F#1\endcsname+\tabcolsep}

\def\cAlignHack{\rightskip\@flushglue\leftskip\@flushglue\parindent\z@\parfillskip\z@skip}
\def\rAlignHack{\rightskip\z@skip\leftskip\@flushglue \parindent\z@\parfillskip\z@skip}

\@ifundefined{etal}{}{}

\usepackage{ifxetex}
\ifxetex\else\if@twocolumn\@ifpackageloaded{stfloats}{}{\usepackage{dblfloatfix}}\fi\fi

\AtBeginDocument{
\expandafter\ifx\csname eqalign\endcsname\relax
\def\eqalign#1{\null\vcenter{\def\\{\cr}\openup\jot\m@th
  \ialign{\strut$\displaystyle{##}$\hfil&$\displaystyle{{}##}$\hfil
      \crcr#1\crcr}}\,}
\fi
}

\AtBeginDocument{%
  \@ifpackageloaded{endfloat}%
   {\renewcommand\efloat@iwrite[1]{\immediate\expandafter\protected@write\csname efloat@post#1\endcsname{}}}{\newif\ifefloat@tables}%
}%

\def\BreakURLText#1{\@tfor\brk@tempa:=#1\do{\brk@tempa\hskip0pt}}
\let\lt=<
\let\gt=>
\def\processVert{\ifmmode|\else\textbar\fi}

\@ifundefined{subparagraph}{
\def\subparagraph{\@startsection{paragraph}{5}{2\parindent}{0ex plus 0.1ex minus 0.1ex}%
{0ex}{\normalfont\small\itshape}}%
}{}

\newcommand\role[1]{\unskip}
\newcommand\aucollab[1]{\unskip}
  
\@ifundefined{tsGraphicsScaleX}{\gdef\tsGraphicsScaleX{1}}{}
\@ifundefined{tsGraphicsScaleY}{\gdef\tsGraphicsScaleY{.9}}{}
\def\checkGraphicsWidth{\ifdim\Gin@nat@width>\linewidth
	\tsGraphicsScaleX\linewidth\else\Gin@nat@width\fi}

\def\checkGraphicsHeight{\ifdim\Gin@nat@height>.9\textheight
	\tsGraphicsScaleY\textheight\else\Gin@nat@height\fi}

\def\fixFloatSize#1{}
\let\ts@includegraphics\includegraphics

\def\inlinegraphic[#1]#2{{\edef\@tempa{#1}\edef\baseline@shift{\ifx\@tempa\@empty0\else#1\fi}\edef\tempZ{\the\numexpr(\numexpr(\baseline@shift*\f@size/100))}\protect\raisebox{\tempZ pt}{\ts@includegraphics{#2}}}}

\AtBeginDocument{\def\includegraphics{\@ifnextchar[{\ts@includegraphics}{\ts@includegraphics[width=\checkGraphicsWidth,height=\checkGraphicsHeight,keepaspectratio]}}}

\DeclareMathAlphabet{\mathpzc}{OT1}{pzc}{m}{it}

\def\URL#1#2{\@ifundefined{href}{#2}{\href{#1}{#2}}}

\def\UrlOrds{\do\*\do\-\do\~\do\'\do\"\do\-}%
\g@addto@macro{\UrlBreaks}{\UrlOrds}

\edef\fntEncoding{\f@encoding}

\makeatother

\newif\ifmultipleabstract\multipleabstractfalse%
\usepackage[nomarkers,tablesfirst,nolists]{endfloat}

\def\fixFloatSize#1{}

\begin{document}

\title{Giant magnon spin conductivity approaching the two-dimensional transport regime in ultrathin yttrium iron garnet films}
\author[1,*]{X-Y. Wei}
\author[1]{O. Alves Santos}
\author[1,$^{\dag}$]{C.H. Sumba Lusero}
\author[1,2]{G. E. W. Bauer}
\author[3]{J. Ben Youssef}
\author[1]{B. J. van Wees}

\affil[1]{Physics of Nanodevices, Zernike Institute for Advanced Materials, University of Groningen, Nijenborgh 4, 9747 AG Groningen, The Netherlands}
\affil[2]{WPI-AIMR $\&$ Institute for Materials Research $\&$ CSRN, Tohoku University, Sendai 980-8577, Japan}
\affil[3]{Lab-STICC, CNRS- UMR 6285, Universit\'e de Bretagne Occidentale, 6 Avenue Le Gorgeu, 29238 Brest Cedex 3, France}

\affil[*]{e-mail: x.wei@rug.nl}
\affil[$^{\dag}$]{Current address: Leibniz Institute for Solid State and Materials Research, IFW, 01069 Dresden, Germany}

\maketitle






\textbf{Conductivities are key material parameters that govern various types of transport (electronic charge, spin, heat etc.)
driven by thermodynamic forces. Magnons, the elementary excitations of the
magnetic order, flow under the gradient of a magnon chemical
potential\cite{Cornelissen_2016,control_us,PhysRevX.10.021029} in proportion to a magnon (spin)
conductivity }$\sigma_{m}$\textbf{. The magnetic insulator yttrium iron
garnet (YIG)\ is the material of choice for efficient magnon spin transport. Here we
report an unexpected giant }$\sigma_{m}$\textbf{\ in record-thin YIG films
with thicknesses down to 3.7 nm when the number of occupied two-dimensional (2D) subbands is
reduced from a large number\ to a few, which corresponds to a transition from 3D to 2D magnon transport. We extract a 2D spin conductivity }($\approx1$\thinspace S)\textbf{\ at room temperature, comparable to the (electronic) spin conductivity of the high-mobility two-dimensional
electron gas in GaAs quantum wells at millikelvin temperatures}\textbf{\cite{high_u_2deg}.
Such high conductivities offer unique opportunities to develop low-dissipation magnon-based
spintronic devices.}

The spin current density in metals is the difference of the up- and down-spin charge
current densities measured in A/m$^{2},$ which is driven by a gradient of the
spin chemical potential (often called spin accumulation) $\boldsymbol{\nabla
}\mu_{s}$. The spin conductivity $\sigma$, defined as\ $j_{s}=j_{\uparrow
}-j_{\downarrow}=\sigma_{s}\boldsymbol{\nabla}\mu_{s}/e$, can be expressed in electrical units as S/m. In
a magnetic insulator where charge currents are absent, each magnon carries
angular momentum $\hbar,$ which is equivalent to the spin current in metals
carried by a pair of spin-up ($+\hbar/2$) and spin-down ($-\hbar/2$) electrons that flow in opposite
directions.
A magnon current $j_{m}$ can be defined as its number current times electron charge
$e$. In magnetic insulator-based spintronic devices, magnon spin currents are
injected, detected, and modulated by microwave striplines or electric contacts
made from a heavy metal for charge-spin conversion
\cite{insulatronics,Cornelissen2015,antiferro,chumak_transistor,PhysRevLett.120.097702}. The corresponding spin conductivity, magnon conductivity $\sigma_{m}$, is the current
density divided by the gradient of the magnon chemical potential. The unit of the magnon conductivity in $j_{m}=\sigma_{m}%
\boldsymbol{\nabla}\mu_{m}/e$, where $\mu_{m}$ is the magnon chemical
potential, is then the same as that of electrons in a
metal\cite{Cornelissen_2016}. The value of $\sigma_{m}=4\times10^{5}\,$S/m in a 210 nm
thick YIG film at room temperature\cite{Cornelissen2015} corresponds to the
electronic conductivity of bad metals.

The high magnetic and acoustic quality of magnetic insulators make them the
ideal material for all-magnon logical circuits and magnon-based quantum
information\cite{roadmap_magnonics,roadmap_computing}. An example of recent
progress in magnon-based computing is an integrated magnonic half-adder based
on 350 nm wide wave guides make from 85-nm-thick YIG films\cite{half_adder}.
However, these devices operate with coherent magnons ($\sim\mathrm{GHz}$)
excited by narrow microwave striplines which can not be integrated into an all-electrical circuit. Therefore, it is attractive to inject
magnons electrically\cite{all_elect}, but those are mainly thermal ($\sim\mathrm{Thz}$) and
scatter much stronger at phonons. Also, scalability to smaller structure sizes,  essential for future high-performance processing units, requires micro and
nanofabrication in all dimensions.
The first step is the growth of films of a few or even a single unit
cell. Previously, magnon transport was reported in transistor structures
on films down to about 10 nm, which shows that ultrathin films can maintain high
quality and display intriguing non-linear magnon
effects\cite{1812.01334,PhysRevB.103.214425}. However, the scattering by
surface roughness is expected to be increase in even thinner
films\cite{rough_surface}. This could be an obstacle for magnon spin
transport in ultrathin YIG films that hinders observation of a transition from
three dimensional magnons to two dimensional magnon gas when the thermal
wavelength $\lambda_{\mathrm{thermal}}=2\pi \sqrt{\hbar\gamma D /(k_{B}T)}$ ($\sim$3 nm at room temperature) approaches the thickness of the films $t_{\mathrm{YIG}}$, where $\hbar$ is the reduced Planck constant, $\gamma$ is the gyromagnetic ratio, $D$ is spin wave stiffness and $k_{B}$ is Boltzmann constant.

Here we report measurements of the magnon conductivity of YIG films with
thicknesses down to 3.7 nm. Much to our surprise, the magnon transport turns
out to be strongly enhanced in the ultrathin regime. We report a drastical
increase in magnon conductivity of up to $\sigma_{m}=1.6\times10^{8}$ S/m at
room temperature that even exceeds the electronic spin conductivity of high-purity
copper. This increase is intimately connected to the small number of occupied
subbands and apparent domination by the lowest subband in our films. These
results can importantly boost the performance of magnon-based information
technology\cite{doi:10.1063/5.0020277,roadmap_magnonics}.

\begin{figure}[tbh]
\vspace{0pt} \includegraphics[width=1\linewidth]{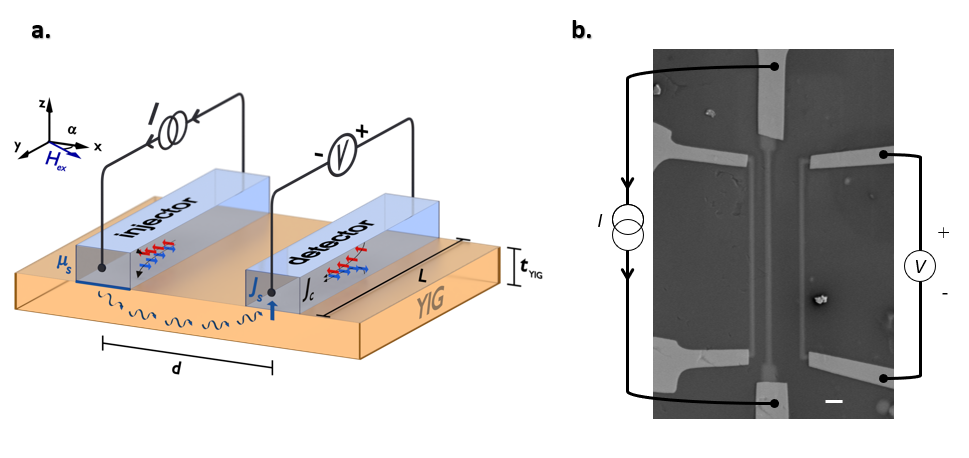}
\vspace{-10pt}\caption{Device layout\textbf{:} \textbf{a)} Schematic
representation of the experimental geometry. Two Pt strips deposited on top of
YIG serve as magnon detector and injector via the direct and inverse spin hall
effect. A low-frequency ac current with rms value of $I_{\mathrm{ac}}$ through
the left Pt strip injects magnons. The center-to-center distance of the
injector and the detector is $d$ and the length of the injector/detector is
$L$. A spin accumulation $\mu_{s}$ is formed at Pt$|$YIG interface due to the
SHE when a charge current passes through the injector and excites magnon
non-equilibrium underneath the injector. The diffusive magnons are absorbed at
the drain, which induce a spin current density $j_{s}$. Using a lock-in technique, the
first harmonic voltage is measured simultaneously by the right Pt strip, i.e.
a magnon detector. $\alpha$ is the angle of external magnetic field
$H_{\mathrm{ex}}$. \textbf{b}) SEM image of the geometry. The parallel
vertical lines are the platinum injector and detector, and they are contacted
by gold leads. Current and voltage connections are indicated schematically.
The scale bar represents 2\thinspace$\mathrm{\mu}$m.}%
\label{fig:device}%
\end{figure}

We employ a non-local configuration\cite{Cornelissen2015} (Figure \ref{fig:device}a) of two Pt thin film
strips with length $L$ at a distance $d$ on top of YIG films grown on gallium
gadolinium garnet (GGG) by liquid phase epitaxy. An electric charge current
$I$ through the injector generates a transverse spin current due to the spin
Hall effect\cite{RevModPhys.87.1213}, resulting in a spin accumulation
$\mu_{s}$ in Pt at the interface to YIG. The injector-conversion coefficient
$\eta_{\mathrm{inj}}=\mu_{s}/(eI)$ depends on the properties and dimensions of
the Pt strip as explained in the Section I of the supplementary information (SI Section I). The effective
interface spin conductance results from the exchange interaction across the
interface and produces a magnon chemical potential $\mu_{m}$ on the
YIG side of the interface that acts as a \emph{magnon source}, where $\mu
_{m}\approx\mu_{s}$ since the interface spin resistance can
be ignored (see SI Section III). The detector electrode is a\emph{\ magnon drain} that absorbs
magnons and converts them into a spin current $j_{s}$ entering the Pt
detector electrode. The inverse spin Hall effect generates a detector voltage
$V_{\mathrm{nl}}$ with detector-conversion coefficient $\eta_{\mathrm{det}%
}=V_{\mathrm{nl}}/j_{s}^{\mathrm{det}}$. By reciprocity, $\eta_{\mathrm{inj}%
}=\eta_{\mathrm{det}}$ when injector and detector contacts have the same
properties (see SI Section I for details). Since the signal scales with $L$, a
normalized non-local resistance can be defined as $R_{\mathrm{nl}%
}=V_{\mathrm{nl}}/\left(  IL\right)  $. The magnon conductance follows from
the measured non-local resistance
\begin{equation}
{G_{m}}=\frac{1}{{\eta}_{\mathrm{inj}}{\eta_{\mathrm{det}}}}\frac
{V_{\mathrm{nl}}}{I}=\frac{R_{\mathrm{nl}}L}{{\eta}_{\mathrm{inj}}%
{\eta_{\mathrm{det}}}}.\label{non_local_conductance}%
\end{equation}
The magnon conductivity $\sigma_{m}$ as a function of the thickness
$t_{\mathrm{YIG}}$ of the YIG films in Figure 1 then follows from the magnon spin
conductance
\begin{equation}
\sigma_{\mathrm{m}}=\frac{G_{\mathrm{m}}d}{t_{\mathrm{YIG}}L}%
.\label{non_local_conductivity}%
\end{equation}
For the films with $t_{\mathrm{YIG}}$ much smaller than the magnon relaxation
length $\lambda_{m}$ as well as the lateral device dimension, $\mu_{m}$ can be considered constant in the $z$ direction
. Therefore, we use
following equation to describe magnon
diffusion\cite{Cornelissen2015,PhysRevB.67.052409}
\begin{equation}
R_{\mathrm{nl}}={\frac{\sigma_{m}t_{\mathrm{YIG}}\eta_{\mathrm{inj}}%
\eta_{\mathrm{det}}}{\lambda_{m}}}\mathrm{csch}\frac{d}{\lambda_{m}%
}{\rightarrow}\left\{
\begin{array}
[c]{c}%
\frac{\sigma_{m}t_{\mathrm{YIG}}\eta_{\mathrm{inj}}\eta_{\mathrm{det}}}{d}\\
\frac{2\sigma_{m}t_{\mathrm{YIG}}\eta_{\mathrm{inj}}\eta_{\mathrm{det}}%
}{\lambda_{m}}\exp(-\frac{d}{\lambda_{m}})
\end{array}
\right.  \text{for }%
\begin{array}
[c]{c}%
d\ll\lambda_{m}\\
d\gg\lambda_{m}%
\end{array}
.\label{1ddiffusion}%
\end{equation}
When the spacing $d$ is smaller than $\lambda_{m}$, it is the Ohmic regime in
which the magnons are conserved, $R_{\mathrm{nl}}(G_{m})\sim d^{-1}.$
Otherwise, the signal decays exponentially as a function of distance due to
magnon relaxation.

\begin{figure}[tbh]
\vspace{0pt} \includegraphics[width=1\linewidth]{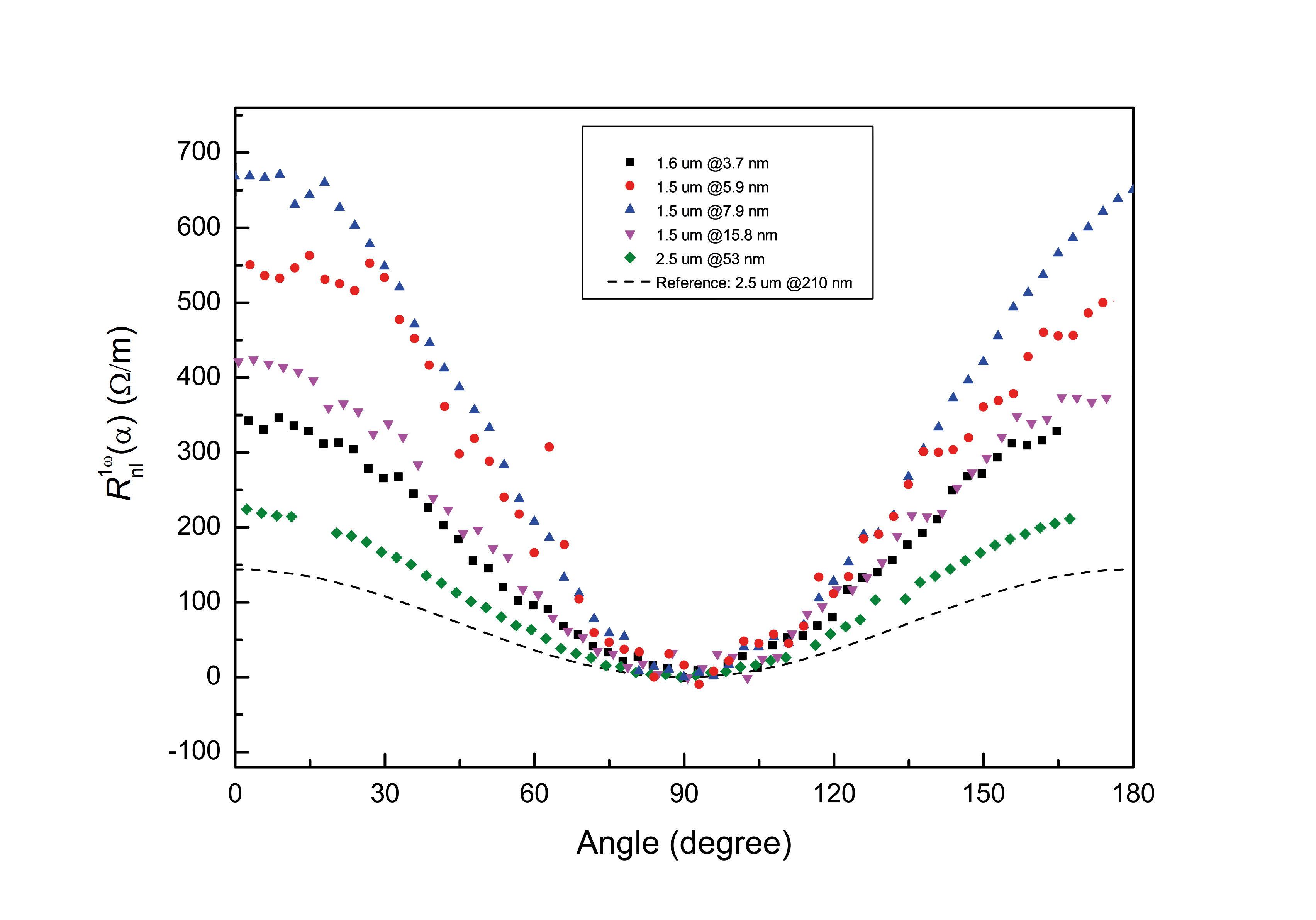}
\vspace{-10pt}\caption{Dependence of non-local resistance on magnetization
direction on ultrathin YIG films, for short center-to-center distances between the
injector and the detector. The offset
$R_{0}^{1\omega}$ in Eq.\ref{angle} has been subtracted. This shows
$R_{\mathrm{nl}}^{1\omega}$ increases with decreasing thickness. Comparing with
the reference from Cornelissen et al.\cite{Cornelissen2015},
$R_{\mathrm{nl}}^{1\omega}$ significantly increases in ultrathin films.}%
\label{fig:e_measure}%
\end{figure}

\begin{figure}[ptb]
\vspace{0pt} \includegraphics[width=1\linewidth]{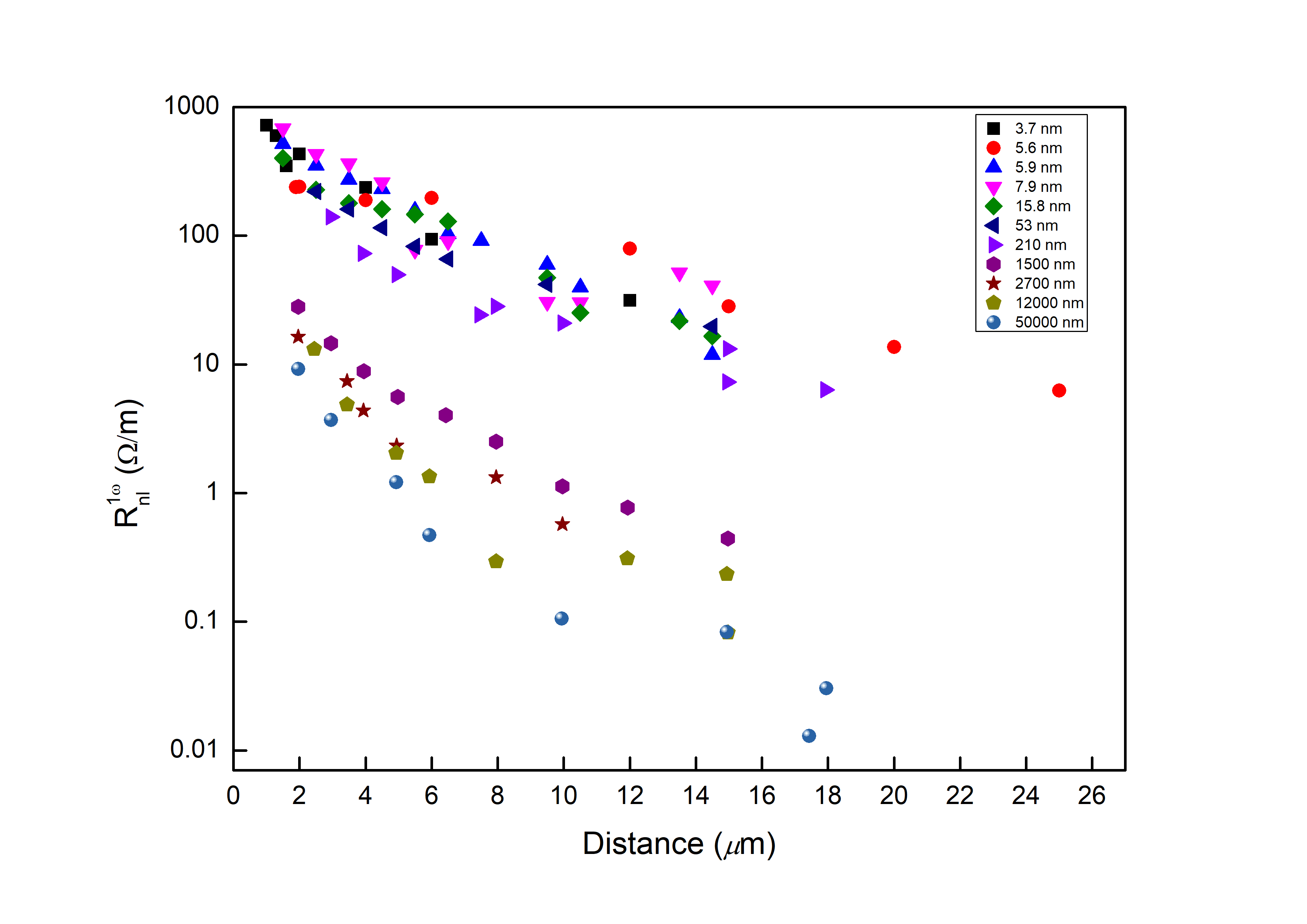}
\vspace{-10pt}\caption{Non-local resistance as a function of injector-detector
distance of the samples of series A and $t_{\mathrm{YIG}}$ varying from 3.7 nm
to 50000 nm. The width of injector/detector is 400 nm. The results for
$t_{\text{YIG}} \geq$ 210 nm are adopted from Cornelissen et
al.\cite{Cornelissen2015} and Shan et al.\cite{PhysRevB.94.174437}.}%
\label{fig:e_together_400}%
\end{figure}

\begin{figure}[ptb]
\center{\vspace{0pt} \includegraphics[width=0.8\linewidth]{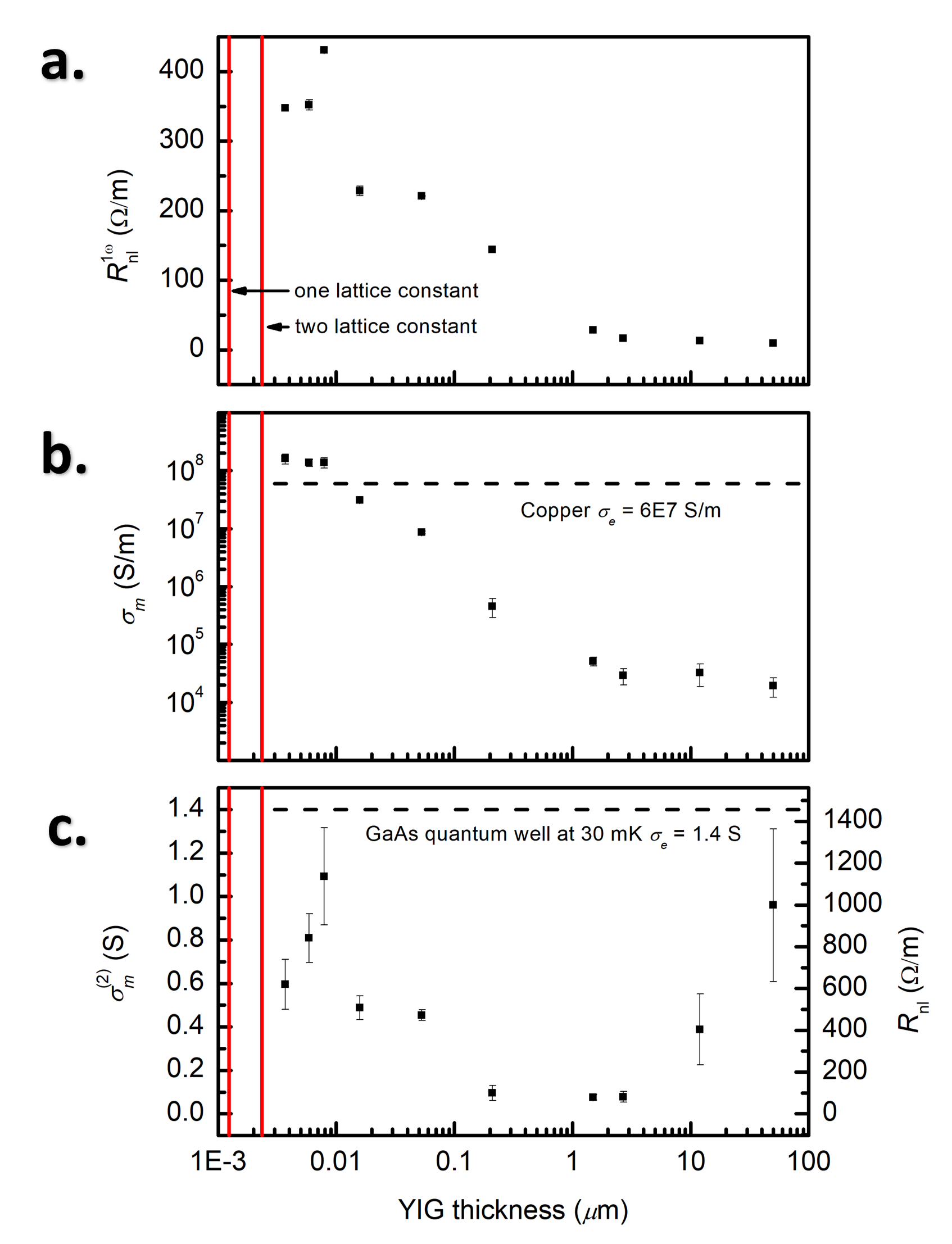}}
\vspace{-10pt}\caption{\textbf{a)} The non-local resistance $R_{\mathrm{nl}}^{1\omega}$ at
$d=2.5\ \mu$m as a function of $t_{\mathrm{YIG}}$. The results for
$t_{\mathrm{YIG}}\geq210\,$nm are adopted from Cornelissen et
al.\cite{Cornelissen2015} and Shan et al.\cite{PhysRevB.94.174437}. \textbf{b) }Thickness dependence of the magnon conductivity
$\sigma_{m}$ obtained by the best fit for different distances with statistical
error bars. \textbf{c)} Thickness dependence of the 2-dimension spin
conductance $\sigma_{m}^{\left(  2\right)  }$ and the non-local resistance $R_{\mathrm{nl}}$ from the
simulation, values based on the best fit for the
magnon conductivity. The saturation at $t_{\mathsf{YIG}}\rightarrow0$
indicates that the film approaches the two-dimensional regime in ultrathin
limit. The obtained error bar here means the range of the best fitting results for the
non-local resistance we can get from the FEM simulations (see SI Section II).}%
\label{fig:YIG_sigma}%
\end{figure}

We measure $R_{\mathrm{nl}}$ at room temperature as a function of an external
in-plane magnetic field $\mathbf{H}_{\mathrm{ex}}$ with $\left\vert
\mathbf{H}_{\mathrm{ex}}\right\vert =50$\thinspace mT, which we rotate in the
plane (Figure \ref{fig:device}a). We modulate the AC current $I$ by a low frequency (18\thinspace Hz) and
detect the first/second harmonic signal $V_{\mathrm{nl}}(\omega)$%
/$V_{\mathrm{nl}}(2\omega)$ by lock-in amplifiers (see Methods).
$V_{\mathrm{nl}}(2\omega)$ depends on the spin Seebeck generation and
diffusion of magnons under an inhomogeneous temperature profile, which renders
interpretation difficult\cite{obser_sse,PhysRevB.101.184420} (see SI Section V). Therefore, we focus on $V_{\mathrm{nl}%
}(\omega)$ that follows the formula
\begin{equation}
R_{\mathrm{nl}}^{1\omega}(\alpha)={R_{\mathrm{nl}}^{1\omega}}\cos^{2}%
\alpha+{R_{0}^{1\omega}},\label{angle}%
\end{equation}
where $R_{0}^{1\omega}$ is an offset resistance (see Methods) and $\alpha$ is
the angle of $\mathbf{H}_{\mathrm{ex}}$ with the $x$-axis. In Figure. \ref{fig:e_measure}, the the angle-dependent measurements in various thickness YIG films show that $R_{\mathrm{nl}}^{1\omega}$ becomes four times larger when the film is over fifty times thinner from 210 nm to 3.7 nm. We also observe a
strongly increased non-local signal in ultrathin films in Figure
\ref{fig:e_together_400} as a function of contact separation for a wide range
of $t_{\mathrm{YIG}}$ including results on ultrathin YIG films for
400\thinspace nm wide Pt strips and for thicker films $t_{\mathrm{YIG}}%
\geq210$\thinspace nm\cite{Cornelissen2015,PhysRevB.94.174437}. Figure
\ref{fig:YIG_sigma}a emphasizes the dramatic enhancement of $R_{\mathrm{nl}%
}^{1\omega}$ for the thinnest films down to $t_{\mathrm{YIG}}=3.7$ nm and
fixed $d=2.5\,\mathrm{\mu}$m, which can be attributed to the $t_{\mathrm{YIG}%
}$ dependence of $\sigma_{m}$ because $\lambda_{m}>2.5$ $\mathrm{\mu}$m for
all thicknesses (see SI Section IV for details). $R_{\mathrm{nl}}^{1\omega}$
\emph{increases} with \emph{decreasing} thickness and saturates for both the thinnest and
thickest films.

A finite-element model\cite{Cornelissen_2016} can simulate the depth ($z$)
dependence of $\mu_{m}$ when $t_{\mathrm{YIG}}>\lambda_{m}$ (see SI Section I for
details). This leads to a limiting $\sigma_{m}\rightarrow3\times10^{4}$ S/m in
Figure \ref{fig:YIG_sigma}b for thicker films, which represents
the bulk value. The simulated $R{_{\mathrm{nl}}}$ values for
$d=2.5\,\mathrm{\mu}$m in Figure \ref{fig:YIG_sigma}c have been fitted to
$R_{\mathrm{nl}}^{1\omega}$ in Figure \ref{fig:YIG_sigma}a by conductivities
that are strongly enhanced in the regime $t_{\mathrm{YIG}}<\lambda_{m}$. For
$t_{\mathrm{YIG}}=3.7$ nm, the magnon conductivity $\sigma_{m}=1.6\times
10^{8}$ S/m is \emph{four orders of magnitude larger} compared to the bulk value,
exceeding the electronic conductivity of pure metals such as copper with
$\sigma_{e}=6\times10^{7}$ S/m \cite{copper}. The observed saturation at
$t_{\mathrm{YIG}}\rightarrow0$ appears to reflect an increased role of surface
roughness scattering that we do not model explicitly.

A magnon conductivity that diverges for $t_{\mathrm{YIG}}\rightarrow0$ like
$\sigma_{m}\sim\sigma_{m}^{\mathrm{2D}}t_{\mathrm{YIG}}^{-1}$ simply suggests two-dimensional transport. 
In Figure \ref{fig:YIG_sigma}c, it shows that $\sigma_{m}^{\mathrm{2D}}$ saturates for $t_{\mathrm{YIG}}<10$ nm, i.e. higher 2D subbands do not contribute significantly even though they are still populated (see below). Extrapolation to zero thickness leads to $\sigma_{m}^{\mathrm{2D}%
}\approx1\,$S. This value at room temperature is comparable to that of the high-mobility
two-dimensional electron gas at millikelvin temperatures, which is $\sigma_{\mathrm{e}%
}^{\mathrm{2D}}\approx1.4\,$S in GaAs quantum wells\cite{high_u_2deg}.

%

The magnons propagate in the plane with wave vector $\mathbf{k}$ and form
perpendicular standing spin waves (PSSW) in $z$ direction labeled by an
integer $n$. The exchange interaction scales like $\sim k^{2}$ and dominates
the magnon dispersion $\varepsilon_{nk}$ at thermal energies ($\approx k_{B}%
T$) with small magnetodipolar corrections. A magnon with energy $\varepsilon
_{nk}=\hbar\gamma D\left(  k^{2}+\left(  n\pi/t_{\mathrm{YIG}}\right)
^{2}\right)  $ contributes to the conduction proportional to its thermal
occupation $N_{nk}=1/\left\{  \exp\left[  \varepsilon_{nk}/\left(
k_{B}T\right)  \right]  -1\right\}  $.\ For YIG $\gamma/2\pi
=28$\thinspace GHz/T and the spin wave stiffness\cite{Klingler_2014}
$D=5\times10^{-17}$\thinspace T$\text{m}^{2}$.
The highest occupied subband $n$ defined as
\begin{equation}
n=\mathrm{int} \left( \frac{t_{\mathrm{YIG}}}{\pi}\sqrt{\frac{k_{B}T}{\hbar\gamma D}} \right)
\label{eq:n_max}%
\end{equation}
at $\varepsilon_{n0}<k_{B}T$ as a function of thickness\cite{stamps2014solid},
where int($x$) is the greatest integer no more than $x$. For $t_{\mathrm{YIG}}=3.7$
nm, only three approximately 2D subbands are occupied at room temperature.

The simplest model for the magnon conductivity in $\nu$ (=2, 3) dimensions follows
from the Boltzmann equation with a constant relaxation time $\tau$
\begin{equation}
\sigma_{m}^{\left(  \nu\right)  }=\frac{e^{2}\tau^{\left(  \nu\right)  }%
}{k_{B}T}\int\frac{d\mathbf{k}}{\left(  2\pi\right)  ^{\nu}}\left(
\frac{\partial\varepsilon_{k}}{\hbar\partial k_{z}}\right)  ^{2}%
\frac{e^{\varepsilon_{k}/\left(  k_{B}T\right)  }}{(e^{\varepsilon_{k}/\left(
k_{B}T\right)  }-1)^{2}}%
\end{equation}
where $\varepsilon_{k}=\hbar\gamma Dk^{2}$. Magnetic freeze-out experiments
show that the contributions from the low-frequency magnons ($\sim\mathrm{GHz}%
$) is significant even at room temperature, presumably reflecting low
mobilities of thermal exchange magnons
\cite{PhysRevB.92.064413,PhysRevB.92.054436,PhysRevB.100.134402}. This can be
represented by a high momentum cut-off $K_{\infty}\sim1/\mathrm{nm}$ at magnon
frequencies $\varepsilon_{\infty}/\hbar\sim\mathrm{THz}$. In the high
temperature limit $k_{B}T\gg\varepsilon_{k}$ the conductivities do not depend
on $\gamma D:$
\begin{equation}
\sigma_{m}^{\left(  3\right)  }=\frac{2 e^{2} k_{B}T\tau^{\left(  3\right)  }}%
{3\hbar^{2}\pi^{2}}K_{\infty},%
\label{sigma_3d}%
\end{equation}%
\begin{equation}
\sigma_{m}^{\left(  2\right)  }=\frac{e^{2} k_{B}T\tau^{\left(  2\right)  }}%
{\pi\hbar^{2}}\mathrm{log}\frac{K_{\infty}}{K_{0}}%
,\label{sigma_2d}%
\end{equation}
where $K_{0}$ is a low momentum cutoff by the magnon gap of $\varepsilon
_{0}/\hbar\sim\mathrm{GHz}$. By equating these equations with the experimental
results $\sigma_{m}^{\left(  3\right)  }\approx3\times10^{4}\,$S/m and the
present $\sigma_{m}^{\left(  2\right)  }\approx1\,$S and using the scattering
times as adjustable parameters, we arrive at $\tau^{\left(  3\right)  }%
\approx40\,$fs and $\tau^{\left(  2\right)  }\approx0.1\,$ns. The short
scattering time in three dimensions can be explained by highly efficient
magnon-phonon scattering at room temperature\cite{Cornelissen_2016}. While the
high-momentum cut-off plays an important role in 3 dimensions, the near
independence of $\sigma_{m}^{\left(  2\right)  }$ emphasizes the importance of
the near band-gap excitations for transport in two dimensions. Coherent
magnons excited at GHz frequencies can propagate over cm's in spite of their
small group velocity because they scatter only weakly at phonons\cite{PhysRevB.99.184442}. Their contribution
has a much larger effect on transport in ultrathin films than in the bulk,
which is consistent with the magnetic field and temperature dependence
reported in the SI. The estimated scattering time of $\tau^{\left(  2\right)
}=0.1\,$ns may be limited by the film roughness scattering. The precise
mechanism can be elucidated only by more extensive experimental and
theoretical studies of the temperature and field dependence.


While magnon-based devices do not suffer from Joule heating, magnon
transport is not dissipationless\cite{PhysRevB.96.100406,Cornelissen2015} even for transport on length scales shorter than the magnon relaxation length where magnons are conserved. The
observed giant magnon conductivity is therefore excellent news, implying low dissipation from magnon-phonon scattering even at room temperature. Ultra-thin films can
therefore be driven with relative ease into the non-linear regime in e.g.
magnon spin transistors\cite{1812.01334,PhysRevB.103.214425}, facilitating
electrically-induced magnon Bose-Einstein condensation and magnon spin
superfluidity\cite{PhysRevLett.108.246601,magnon_bec_micro,spin_current_bec}.
The robustness of the transport of the magnetic order for thin films of close to the monolayer
thickness should allow magnon transport in nanostructures such as constrictions, wires and dots with
feature sizes of a few nanometer without loss of magnetic functionality.

\section*{Methods}

\subsection{Fabrication}

The YIG films are grown on $\text{Gd}_{3}$$\text{Ga}_{5}$$\text{O}_{12}$
(GGG) substrates by liquid-phase epitaxy (LPE) at the Universit\'{e} de
Bretagne Occidentale in Brest, France, with thicknesses from 3.7 nm to 53 nm
The effective magnetization $M_{\mathrm{eff}}$ ($H_{\mathrm{k}}-4\pi M_{\mathrm{s}}$) and the magnetic relaxation (intrinsic damping parameter $\alpha$ and extrinsic inhomogenous linewidth $\Delta H_{\mathrm{in}}$) are determined by broadband ferromagnetic resonance (FMR) in the frequency range 2-40 GHz (see SI Section IV). The device patterns are written by three e-beam lithography steps,
each followed by a standard deposition and lift-off procedure. The first step
produces a Ti/Au marker pattern, used to align the subsequent steps. The
second step defines the platinum injector and detector strips, as deposited by
dc sputtering in an Ar+ plasma at an argon pressure with thickness
$\symbol{126}8$\thinspace nm for all devices. The third step defines 5/75 nm
Ti/Au leads and bonding pads, deposited by e-beam evaporation. Devices have an
injector/detector length $L=30/25$ $\mathrm{\mu}$m and the strip widths $W$
are 400 nm for series A and 100 nm for series B. The experimental results in
main text are obtained from series A. The distance-dependent non-local
resistances for series B can be found in SI Section III.

\subsection{Measurements}

All measurements were carried out by means of three SR830 lock-in amplifiers
using excitation frequency of 18 Hz. The lock-in amplifiers are set up to
measure the first and second harmonic responses of the sample. Current was
sent to the sample using a custom built current source, galvanically isolated
from the rest of the measurement equipment. Voltage measurements were made
using a custom-built pre-amplifier (gain $10^{3}$) and amplified further using
the lock-in systems. The typical excitation currents applied to the samples
are 200 $\mathrm{\mu}$A (RMS) for series A and 20 $\mathrm{\mu}%
$m for series B. The in-plane coercive field of the YIG $B_{c}$ is below
10\thinspace mT for all YIG samples, and we apply an external field to orient
the magnetization using a physical property measurement system (PPMS). The
samples are mounted on a rotatable sample holder with stepper motor. All
experimental data in the main text have been collected at 300 K (room
temperature) at an applied magnetic field of 50 mT.

\subsection{Simulations}

Our finite-element model implements the magnon diffusion equation in insulators in order to simulate transport of electrically injected magnons. We carried out the
simulations by COMSOL MULTIPHYSICS (version 5.4) software package with
technical details in the SI Section I.

\section*{Acknowledgements}

We acknowledge the helpful discussion with J. Shan and T. Yu. We acknowledge the technical support from J.
G. Holstein, H. de Vries, H. Adema, T. Schouten and A. Joshua. This work
is part of the research programme "Skyrmionics" with project number 170, which is financed by
the Dutch Research Council (NWO). The support by NanoLab NL and the Spinoza Prize awarded in 2016 to B. J. van Wees by NWO is also gratefully
acknowledged. G.B. was supported by JSPS Kakenhi Grant 19H00645.

\section*{Author contributions}

B.J.v.W. and X.W. conceived the experiments. X.W. designed and carried out the experiments, with help from O.A.S. J.B.Y. supplied the YIG samples used in the fabrication of devices. X.W., O.A.S., C.H.S.L., G.E.W.B. and B.J.v.W. were involved in the analysis. X.W. wrote the paper with O.A.S., G.E.W.B. and B.J.v.W. All authors commented on the manuscript.

\section*{References}

\bibliographystyle{plain}

\begin{thebibliography}{10}
\expandafter\ifx\csname url\endcsname\relax
  \def\url#1{\texttt{#1}}\fi
\expandafter\ifx\csname urlprefix\endcsname\relax\def\urlprefix{URL }\fi
\providecommand{\bibinfo}[2]{#2}
\providecommand{\eprint}[2][]{\url{#2}}

\bibitem{Cornelissen_2016}
\bibinfo{author}{Cornelissen, L.~J.}, \bibinfo{author}{Peters, K. J.~H.},
  \bibinfo{author}{Bauer, G. E.~W.}, \bibinfo{author}{Duine, R.~A.} \&
  \bibinfo{author}{van Wees, B.~J.}
\newblock \bibinfo{title}{Magnon spin transport driven by the magnon chemical
  potential in a magnetic insulator}.
\newblock \emph{\bibinfo{journal}{Phys. Rev. B}} \textbf{\bibinfo{volume}{94}},
  \bibinfo{pages}{014412} (\bibinfo{year}{2016}).

\bibitem{control_us}
\bibinfo{author}{Chunhui, D.} \emph{et~al.}
\newblock \bibinfo{title}{Control and local measurement of the spin chemical
  potential in a magnetic insulator}.
\newblock \emph{\bibinfo{journal}{Science}} \textbf{\bibinfo{volume}{357}},
  \bibinfo{pages}{195--198} (\bibinfo{year}{2017}).

\bibitem{PhysRevX.10.021029}
\bibinfo{author}{Olsson, K.~S.} \emph{et~al.}
\newblock \bibinfo{title}{Pure spin current and magnon chemical potential in a
  nonequilibrium magnetic insulator}.
\newblock \emph{\bibinfo{journal}{Phys. Rev. X}} \textbf{\bibinfo{volume}{10}},
  \bibinfo{pages}{021029} (\bibinfo{year}{2020}).

\bibitem{high_u_2deg}
\bibinfo{author}{Chung, Y.~J.} \emph{et~al.}
\newblock \bibinfo{title}{Ultra-high-quality two-dimensional electron systems}.
\newblock \emph{\bibinfo{journal}{Nature Materials}}
  \textbf{\bibinfo{volume}{20}}, \bibinfo{pages}{632--637}
  (\bibinfo{year}{2021}).

\bibitem{insulatronics}
\bibinfo{author}{Brataas, A.}, \bibinfo{author}{{van Wees}, B.},
  \bibinfo{author}{Klein, O.}, \bibinfo{author}{{de Loubens}, G.} \&
  \bibinfo{author}{Viret, M.}
\newblock \bibinfo{title}{Spin insulatronics}.
\newblock \emph{\bibinfo{journal}{Physics Reports}}
  \textbf{\bibinfo{volume}{885}}, \bibinfo{pages}{1--27}
  (\bibinfo{year}{2020}).

\bibitem{Cornelissen2015}
\bibinfo{author}{Cornelissen, L.~J.}, \bibinfo{author}{Liu, J.},
  \bibinfo{author}{Duine, R.~A.}, \bibinfo{author}{Ben~Youssef, J.} \&
  \bibinfo{author}{van Wees, B.~J.}
\newblock \bibinfo{title}{{Long-distance transport of magnon spin information
  in a magnetic insulator at room temperature}}.
\newblock \emph{\bibinfo{journal}{Nature Physics}}
  \textbf{\bibinfo{volume}{11}}, \bibinfo{pages}{1022--1026}
  (\bibinfo{year}{2015}).

\bibitem{antiferro}
\bibinfo{author}{Lebrun, R.} \emph{et~al.}
\newblock \bibinfo{title}{Tunable long-distance spin transport in a crystalline
  antiferromagnetic iron oxide}.
\newblock \emph{\bibinfo{journal}{Nature}} \textbf{\bibinfo{volume}{561}},
  \bibinfo{pages}{222--225} (\bibinfo{year}{2018}).

\bibitem{chumak_transistor}
\bibinfo{author}{Chumak, A.~V.}, \bibinfo{author}{Serga, A.~A.} \&
  \bibinfo{author}{Hillebrands, B.}
\newblock \bibinfo{title}{Magnon transistor for all-magnon data processing}.
\newblock \emph{\bibinfo{journal}{Nature Communications}}
  \textbf{\bibinfo{volume}{5}}, \bibinfo{pages}{4700} (\bibinfo{year}{2014}).

\bibitem{PhysRevLett.120.097702}
\bibinfo{author}{Cornelissen, L.~J.}, \bibinfo{author}{Liu, J.},
  \bibinfo{author}{van Wees, B.~J.} \& \bibinfo{author}{Duine, R.~A.}
\newblock \bibinfo{title}{Spin-current-controlled modulation of the magnon spin
  conductance in a three-terminal magnon transistor}.
\newblock \emph{\bibinfo{journal}{Phys. Rev. Lett.}}
  \textbf{\bibinfo{volume}{120}}, \bibinfo{pages}{097702}
  (\bibinfo{year}{2018}).

\bibitem{roadmap_magnonics}
\bibinfo{author}{Barman, A.} \emph{et~al.}
\newblock \bibinfo{title}{{The 2021 Magnonics Roadmap}}.
\newblock \emph{\bibinfo{journal}{Journal of Physics: Condensed Matter}}
  \textbf{\bibinfo{volume}{33}}, \bibinfo{pages}{413001}
  (\bibinfo{year}{2021}).

\bibitem{roadmap_computing}
\bibinfo{author}{Chumak, A.} \emph{et~al.}
\newblock \bibinfo{title}{{Roadmap on Spin-Wave Computing. Preprint at:
  https://arxiv.org/abs/2111.00365}}  (\bibinfo{year}{2021}).

\bibitem{half_adder}
\bibinfo{author}{Wang, Q.} \emph{et~al.}
\newblock \bibinfo{title}{A magnonic directional coupler for integrated
  magnonic half-adders}.
\newblock \emph{\bibinfo{journal}{Nature Electronics}}
  \textbf{\bibinfo{volume}{3}}, \bibinfo{pages}{765--774}
  (\bibinfo{year}{2020}).

\bibitem{all_elect}
\bibinfo{author}{Althammer, M.}
\newblock \bibinfo{title}{All-electrical magnon transport experiments in
  magnetically ordered insulators}.
\newblock \emph{\bibinfo{journal}{physica status solidi (RRL) --Rapid Research
  Letters}} \textbf{\bibinfo{volume}{15}}, \bibinfo{pages}{2100130}
  (\bibinfo{year}{2021}).

\bibitem{1812.01334}
\bibinfo{author}{Wimmer, T.} \emph{et~al.}
\newblock \bibinfo{title}{Spin transport in a magnetic insulator with zero
  effective damping}.
\newblock \emph{\bibinfo{journal}{Phys. Rev. Lett.}}
  \textbf{\bibinfo{volume}{123}}, \bibinfo{pages}{257201}
  (\bibinfo{year}{2019}).

\bibitem{PhysRevB.103.214425}
\bibinfo{author}{Liu, J.}, \bibinfo{author}{Wei, X.-Y.},
  \bibinfo{author}{Bauer, G. E.~W.}, \bibinfo{author}{Youssef, J.~B.} \&
  \bibinfo{author}{van Wees, B.~J.}
\newblock \bibinfo{title}{Electrically induced strong modulation of magnon
  transport in ultrathin magnetic insulator films}.
\newblock \emph{\bibinfo{journal}{Phys. Rev. B}}
  \textbf{\bibinfo{volume}{103}}, \bibinfo{pages}{214425}
  (\bibinfo{year}{2021}).

\bibitem{rough_surface}
\bibinfo{author}{Yu, T.}, \bibinfo{author}{Sharma, S.},
  \bibinfo{author}{Blanter, Y.~M.} \& \bibinfo{author}{Bauer, G. E.~W.}
\newblock \bibinfo{title}{Surface dynamics of rough magnetic films}.
\newblock \emph{\bibinfo{journal}{Phys. Rev. B}} \textbf{\bibinfo{volume}{99}},
  \bibinfo{pages}{174402} (\bibinfo{year}{2019}).

\bibitem{doi:10.1063/5.0020277}
\bibinfo{author}{Li, Y.} \emph{et~al.}
\newblock \bibinfo{title}{Hybrid magnonics: Physics, circuits, and applications
  for coherent information processing}.
\newblock \emph{\bibinfo{journal}{Journal of Applied Physics}}
  \textbf{\bibinfo{volume}{128}}, \bibinfo{pages}{130902}
  (\bibinfo{year}{2020}).

\bibitem{RevModPhys.87.1213}
\bibinfo{author}{Sinova, J.}, \bibinfo{author}{Valenzuela, S.~O.},
  \bibinfo{author}{Wunderlich, J.}, \bibinfo{author}{Back, C.~H.} \&
  \bibinfo{author}{Jungwirth, T.}
\newblock \bibinfo{title}{{Spin Hall effects}}.
\newblock \emph{\bibinfo{journal}{Rev. Mod. Phys.}}
  \textbf{\bibinfo{volume}{87}}, \bibinfo{pages}{1213--1260}
  (\bibinfo{year}{2015}).

\bibitem{PhysRevB.67.052409}
\bibinfo{author}{Takahashi, S.} \& \bibinfo{author}{Maekawa, S.}
\newblock \bibinfo{title}{Spin injection and detection in magnetic
  nanostructures}.
\newblock \emph{\bibinfo{journal}{Phys. Rev. B}} \textbf{\bibinfo{volume}{67}},
  \bibinfo{pages}{052409} (\bibinfo{year}{2003}).

\bibitem{obser_sse}
\bibinfo{author}{Uchida, K.} \emph{et~al.}
\newblock \bibinfo{title}{{Observation of the spin Seebeck effect}}.
\newblock \emph{\bibinfo{journal}{Nature}} \textbf{\bibinfo{volume}{455}},
  \bibinfo{pages}{778--781} (\bibinfo{year}{2008}).

\bibitem{PhysRevB.101.184420}
\bibinfo{author}{Gomez-Perez, J.~M.}, \bibinfo{author}{V\'elez, S.},
  \bibinfo{author}{Hueso, L.~E.} \& \bibinfo{author}{Casanova, F.}
\newblock \bibinfo{title}{{Differences in the magnon diffusion length for
  electrically and thermally driven magnon currents in
  ${\mathrm{Y}}_{3}\mathrm{F}{\mathrm{e}}_{5}{\mathrm{O}}_{12}$}}.
\newblock \emph{\bibinfo{journal}{Phys. Rev. B}}
  \textbf{\bibinfo{volume}{101}}, \bibinfo{pages}{184420}
  (\bibinfo{year}{2020}).

\bibitem{PhysRevB.94.174437}
\bibinfo{author}{Shan, J.} \emph{et~al.}
\newblock \bibinfo{title}{Influence of yttrium iron garnet thickness and heater
  opacity on the nonlocal transport of electrically and thermally excited
  magnons}.
\newblock \emph{\bibinfo{journal}{Phys. Rev. B}} \textbf{\bibinfo{volume}{94}},
  \bibinfo{pages}{174437} (\bibinfo{year}{2016}).

\bibitem{copper}
\bibinfo{author}{Laughton, M.~A.} \& \bibinfo{author}{Say, M.~G.}
\newblock \emph{\bibinfo{title}{Electrical engineer's reference book}}
  (\bibinfo{publisher}{Elsevier}, \bibinfo{year}{2013}).

\bibitem{Klingler_2014}
\bibinfo{author}{Klingler, S.} \emph{et~al.}
\newblock \bibinfo{title}{Measurements of the exchange stiffness of {YIG} films
  using broadband ferromagnetic resonance techniques}.
\newblock \emph{\bibinfo{journal}{Journal of Physics D: Applied Physics}}
  \textbf{\bibinfo{volume}{48}}, \bibinfo{pages}{015001}
  (\bibinfo{year}{2014}).

\bibitem{stamps2014solid}
\bibinfo{author}{Stamps, R.} \& \bibinfo{author}{Camley, R.}
\newblock \emph{\bibinfo{title}{{Solid State Physics}}}.
\newblock No. \bibinfo{number}{Volume 65} in \bibinfo{series}{Solid State
  Physics} (\bibinfo{publisher}{Elsevier Science}, \bibinfo{year}{2014}).

\bibitem{PhysRevB.92.064413}
\bibinfo{author}{Kikkawa, T.} \emph{et~al.}
\newblock \bibinfo{title}{{Critical suppression of spin Seebeck effect by
  magnetic fields}}.
\newblock \emph{\bibinfo{journal}{Phys. Rev. B}} \textbf{\bibinfo{volume}{92}},
  \bibinfo{pages}{064413} (\bibinfo{year}{2015}).

\bibitem{PhysRevB.92.054436}
\bibinfo{author}{Jin, H.}, \bibinfo{author}{Boona, S.~R.},
  \bibinfo{author}{Yang, Z.}, \bibinfo{author}{Myers, R.~C.} \&
  \bibinfo{author}{Heremans, J.~P.}
\newblock \bibinfo{title}{{Effect of the magnon dispersion on the longitudinal
  spin Seebeck effect in yttrium iron garnets}}.
\newblock \emph{\bibinfo{journal}{Phys. Rev. B}} \textbf{\bibinfo{volume}{92}},
  \bibinfo{pages}{054436} (\bibinfo{year}{2015}).

\bibitem{PhysRevB.100.134402}
\bibinfo{author}{Jamison, J.~S.} \emph{et~al.}
\newblock \bibinfo{title}{Long lifetime of thermally excited magnons in bulk
  yttrium iron garnet}.
\newblock \emph{\bibinfo{journal}{Phys. Rev. B}}
  \textbf{\bibinfo{volume}{100}}, \bibinfo{pages}{134402}
  (\bibinfo{year}{2019}).

\bibitem{PhysRevB.99.184442}
\bibinfo{author}{Streib, S.}, \bibinfo{author}{Vidal-Silva, N.},
  \bibinfo{author}{Shen, K.} \& \bibinfo{author}{Bauer, G. E.~W.}
\newblock \bibinfo{title}{Magnon-phonon interactions in magnetic insulators}.
\newblock \emph{\bibinfo{journal}{Phys. Rev. B}} \textbf{\bibinfo{volume}{99}},
  \bibinfo{pages}{184442} (\bibinfo{year}{2019}).

\bibitem{PhysRevB.96.100406}
\bibinfo{author}{Man, H.} \emph{et~al.}
\newblock \bibinfo{title}{Direct observation of magnon-phonon coupling in
  yttrium iron garnet}.
\newblock \emph{\bibinfo{journal}{Phys. Rev. B}} \textbf{\bibinfo{volume}{96}},
  \bibinfo{pages}{100406} (\bibinfo{year}{2017}).

\bibitem{PhysRevLett.108.246601}
\bibinfo{author}{Bender, S.~A.}, \bibinfo{author}{Duine, R.~A.} \&
  \bibinfo{author}{Tserkovnyak, Y.}
\newblock \bibinfo{title}{{Electronic Pumping of Quasiequilibrium
  Bose-Einstein-Condensed Magnons}}.
\newblock \emph{\bibinfo{journal}{Phys. Rev. Lett.}}
  \textbf{\bibinfo{volume}{108}}, \bibinfo{pages}{246601}
  (\bibinfo{year}{2012}).

\bibitem{magnon_bec_micro}
\bibinfo{author}{Demokritov, S.~O.} \emph{et~al.}
\newblock \bibinfo{title}{{Bose--Einstein condensation of quasi-equilibrium
  magnons at room temperature under pumping}}.
\newblock \emph{\bibinfo{journal}{Nature}} \textbf{\bibinfo{volume}{443}},
  \bibinfo{pages}{430--433} (\bibinfo{year}{2006}).

\bibitem{spin_current_bec}
\bibinfo{author}{Divinskiy, B.} \emph{et~al.}
\newblock \bibinfo{title}{{Evidence for spin current driven Bose-Einstein
  condensation of magnons}}.
\newblock \emph{\bibinfo{journal}{Nature Communications}}
  \textbf{\bibinfo{volume}{12}}, \bibinfo{pages}{6541} (\bibinfo{year}{2021}).

\end{thebibliography}

\end{document}


\title{Giant magnon spin conductivity approaching the two-dimensional transport regime in ultrathin yttrium iron garnet films}

\author{X-Y. Wei}
\author{O. Alves Santos}
\author{C.H. Sumba Lusero}
\author{G. E. W. Bauer}
\author{J. Ben Youssef}
\author{B. J. van Wees}

\maketitle


\setcounter{equation}{0} \setcounter{figure}{0} \setcounter{table}{0}
\setcounter{page}{1}
\makeatletter\renewcommand{\theequation}{S\arabic{equation}}
\renewcommand{\thefigure}{S\arabic{figure}}
\renewcommand{\bibnumfmt}[1]{[S#1]} \renewcommand{\citenumfont}[1]{S#1}

\section{Finite-element modelling}
\label{section:finite_element_modelling}

\label{section:Finite-element model} We measure magnon transport non-locally
by monitoring the voltage in a Pt detector as a function of current a Pt
injector. The electrical spin accumulation $\mu_{s}$ (expressed in energy), which
is generated by a current $I$ in the injector at the YIG interface reads
\cite{Sinova_2015}
\begin{equation}
\mu_{s}=2eI\theta_{\mathrm{Pt}}\frac{l_{s}}{{\sigma_{e}}{t}{w}}\tanh\frac
{t}{2l_{s}},\label{eq:spinaccu}%
\end{equation}
where $e$, $t$, $w$, $\theta_{\mathrm{Pt}}$, $l_{s}$ and $\sigma_{s}$ are the electronic charge, the Pt
film thickness, width, spin Hall angle, spin relaxation length, and
conductivity of the Pt contact. With parameters in Table \ref{tab:Parameters}
the charge current to spin accumulation conversion coefficient $\eta
_{\mathrm{inj}}=\mu_{s}/(eI)=0.05$\thinspace$\Omega$ when $w=400\,\text{nm}$ and
$0.21\,\Omega$ when $w=100\,$nm.

\begin{figure}[htbp]
\label{fig:sketch}
\includegraphics[width=0.55\textwidth]{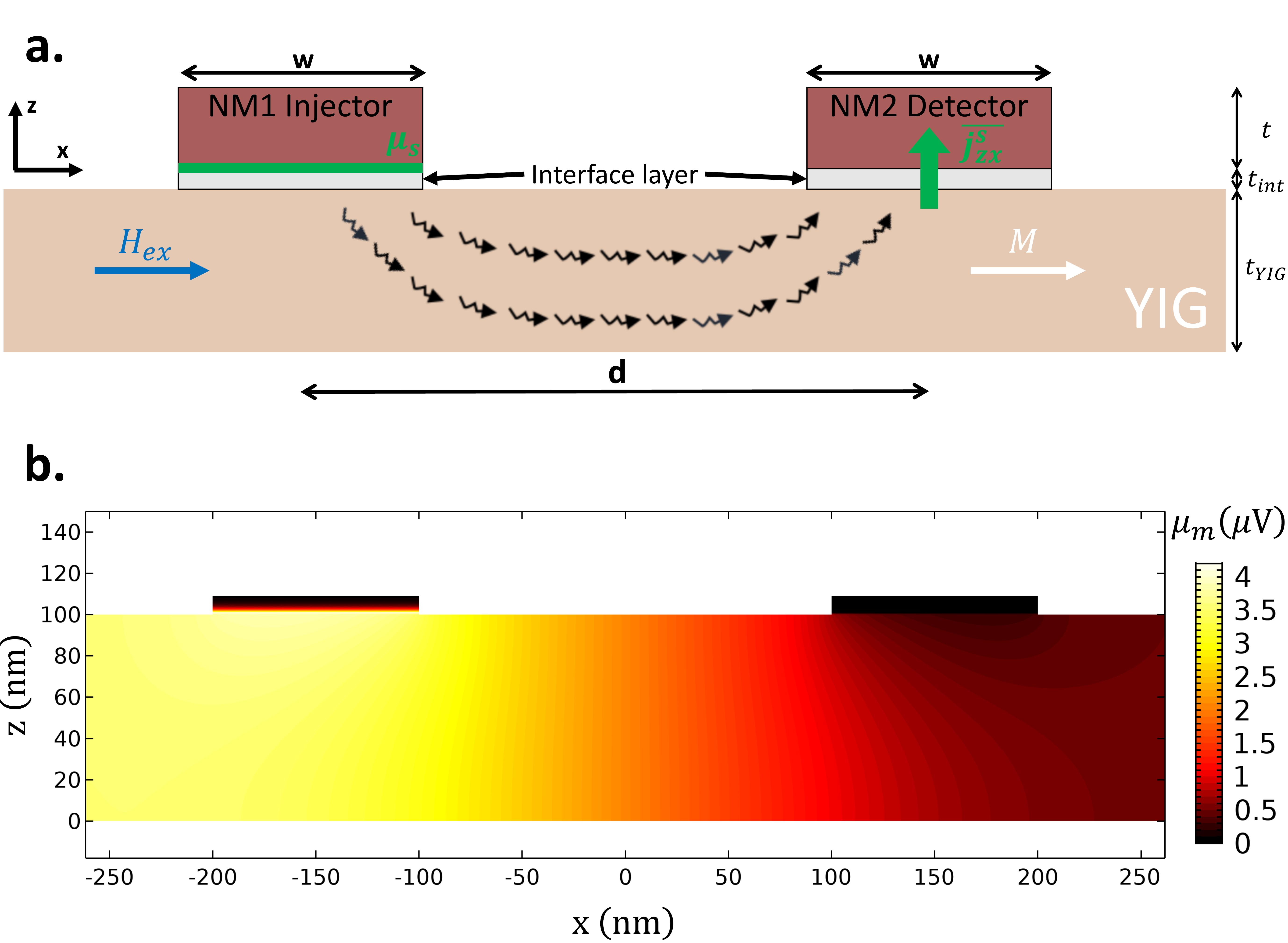}
\caption{\textbf{a)} The sample configuration and its dimensions in the model
simulations.
Eq. \ref{eq:spinaccu} and Eq. \ref{eq:nonlocalob} describe the spin accumulation $\mu_s$ at the interface, used as input of the model, and the voltage build up in the detector, respectively.
\textbf{b)} The magnon chemical
potential profile for a 100$\,$nm thick YIG film with a charge current
$I=20\,\mathrm{\mu}$\textrm{A} and 100$\,$nm wide injector/detector contacts
and parameters in Table. \ref{tab:Parameters}. Magnon absorption by the detector electrode causes the
dark region close to the detector.}%
\end{figure}

We calculate the diffusive magnon spin transport in YIG films numerically by
finite-element method (FEM) \cite{Cornelissen_2016} in the configuration of
Figure \ref{fig:sketch} a) using the COMSOL MULTIPHYSICS (version 5.4)
software package. Since the length of the strips is much longer than the electrode separation, we use a 2D model, where the parameters depend on x and z only.

Ohm's Law for magnon current density $\mathbf{j}_{m}$ in YIG\ is
\begin{equation}
\mathbf{j}_{m}=-{\sigma_{m}}\boldsymbol{\nabla}{\mu_{m}}/e.
\label{eq:diffusive1}%
\end{equation}
where $\boldsymbol{\nabla}=\hat{\boldsymbol{x}}\partial_{x}+\hat
{\boldsymbol{z}}\partial_{z}$ and $\sigma_{m}$ is the magnon spin
conductivity. The local magnon current is proportional
to $\nabla\mu_{m}$. The diffusion equation
\begin{equation}
\nabla^{2}\mu_{m}=\frac{\mu_{m}}{l_{m}^{2}}, \label{eq:diffusive2}%
\end{equation}
governs the magnon chemical potential $\mu_{m},$ where $l_{m}$ is the magnon
relaxation length. We use the zero-current boundary condition $\left(
\boldsymbol{\nabla}\cdot\boldsymbol{n}\right)  \mu_{m}=0$ for the bottom as
well as top surface that is not covered by Pt, where $\boldsymbol{n}$ is the
surface normal.

\begin{table}[ptb]
\caption{Parameters used to obtain the results in Figure \ref{fig:sketch} b)
and Figure \ref{fig:thickness} b) and c).}%
\label{tab:Parameters}
\begin{ruledtabular}
			\begin{tabular}{lcr}
				Parameter & Symbol & Value\\
				\colrule
				Magnon spin conductivity & $\sigma_m$ & $5\times10^{5}$ S/m (for 210 nm thickness YIG \cite{Cornelissen_2016})\\
				Magnon spin relaxation length & $l_m$ & $5\times10^{-6}$ m\\
				Effective spin mixing conductance & $G_s^{eff}$ & $2\times10^{12}$ S/$\textrm{m}^2$\\
				Pt spin Hall angle & $\theta_{\mathrm{Pt}}$ & 0.11\\
				Pt conductivity & $\sigma_e$ & $2\times10^{6}$ S/m\\
				Pt spin relaxation length & $l_s$ & $1.5\times10^{-9}$ m\\
			\end{tabular}
		\end{ruledtabular}
\end{table}

The spin accumulation in the detector generates an magnon transport driving force proportional to $\partial\mu_{s}/\partial z$ that leads to a measurable
voltage by the integral over the cross-section $A=wt$
\begin{equation}
V_{\mathrm{nl}}=\frac{\theta_{\mathrm{Pt}}}{2A}\int\limits_{A}\frac
{\partial\mu_{s}}{\partial z}dA.\label{eq:nonlocalob}%
\end{equation}
The non-local resistance $R_{\mathrm{nl}}=V_{\mathrm{nl}}/\left(  IL\right)  $
(in unit of $\Omega$/m) can be compared with the experimental results. The
detector efficiency is the same as that of the injector, $\eta_{\mathrm{det}%
}=\eta_{\mathrm{inj}}$.




The conversion of spin accumulation in Pt into magnons in YIG is governed by
an effective spin conductance $G_{s}^{\mathrm{eff}},$ which is a certain fraction, $\approx0.06$, of the spin mixing conductance \cite{Althammer_2013,Qui_2013}.
It can be modelled by a hypothetical spacer between Pt and YIG with conductivity $\sigma
_{s}^{\mathrm{int}}=G_{s}^{\mathrm{eff}}t_{\mathrm{int}}$, where $t_{\mathrm{int}}=1\ \text{nm}$ is the spacer layer thickness.

Figure \ref{fig:sketch} b) shows the magnon chemical potential profile
$\mu_{m}$\ in a 100 nm thick YIG under a charge current of $I=20\,\mathrm{\mu
}$A. Injector and detector with $w=100\,$nm/$t=8\,$nm have a center-to-center
distance of 300$\,$nm. Table \ref{tab:Parameters} lists the parameters used to
produce Figure \ref{fig:sketch}b). The Pt detector strip acts as a spin sink,
absorbing magnons and producing a dark YIG region at the bottom of the
detector in Fig. \ref{fig:sketch}b).

\section{YIG film thickness dependence}

\label{section:Fixed_sigma}

\begin{figure}[htbp]
\includegraphics[width=0.9\textwidth]{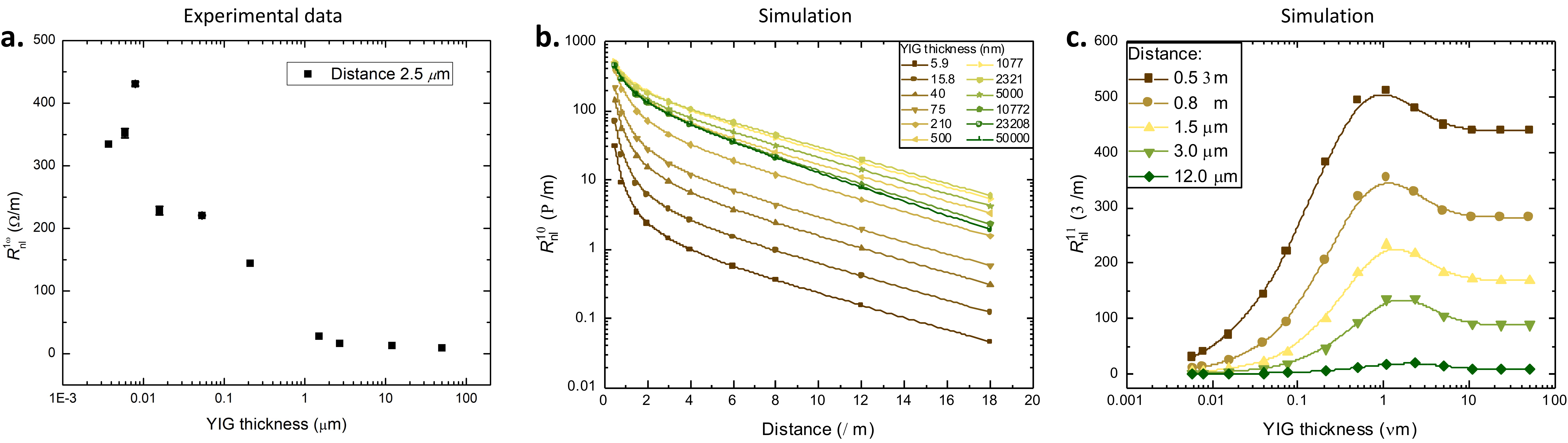}
\caption{\textbf{a)} YIG thickness dependence of the measured non-local signal
$R_{\mathrm{nl}}^{1\omega}$ for a injector/detector distance of
$d=2.5\,\mathrm{\mu m}$. \textbf{b)} Finite element model results with fixed
magnon conductivity $\sigma_{m}=5\times10^{5}$ S/m obtained previously for
$t_{\mathrm{YIG}}=210\,$nm \cite{Cornelissen_2016}. \textbf{c)} $R_{\mathrm{nl}}^{1\omega}$ thickness dependency for different distances, for a fixed value
of magnon spin conductivity.
This shows an opposite trend in comparison with \textbf{a)}.}%
\label{fig:thickness}%
\end{figure}

We measured the non-local resistance in many samples in order to obtain the
YIG film thickness and contact-distance dependence. Figure \ref{fig:thickness} a) shows the experimental results for the non-local resistance for 400 nm wide
injector/detector, \ with center to center distance $d=2.5\,\mathrm{\mu m}$,
and for different YIG thicknesses $t_{\mathrm{YIG}}$. We distinguish two
regimes: when $t_{\mathrm{YIG}}>l_{m}$, $R_{\mathrm{nl}}^{1\omega}$ saturates at a
small value that does not depend on $t_{\mathrm{YIG}}$ anymore. On the other hand,
$R_{\mathrm{nl}}^{1\omega}$ increases with decreasing YIG thickness when
$t_{\mathrm{YIG}}<l_{m}$, where $l_{m}$ is the magnon relaxation length.

Figure \ref{fig:thickness} b) shows that the non-local resistance calculated
with a constant magnon spin conductivity \textit{decreases} when the YIG film
become \textit{thinner}. Figure \ref{fig:thickness} c) shows the results for selected $d$ as a function of $t_{\mathrm{YIG}}\ $that in contrast to experiments, show a completely different thickness dependence.

In Figure \ref{fig:thickness} c), $R_{\mathrm{nl}}^{1\omega}$ saturates above
$t_{\mathrm{YIG}}>l_{m}$ because the magnon current distribution does not change
anymore when increasing $t_{\mathrm{YIG}}$. When $t_{\mathrm{YIG}}<l_{m}$, on
the other hand, the non-local resistance vanishes when $t_{\mathrm{YIG}}=0$,
clearly is opposite to the experimental trend. We have to conclude that
$\sigma_{m}$ dramatically increases the thin-film limit.

We now use $\sigma_{m}$ as a single adjustable parameters that fits the
experimental results.
The numerical simulations are summarized in Figure S3 a) to j).
The values of spin relaxation length used in the simulations, shown in Table \ref{tab:spin_diffusion}, were obtained from the experimental measurements at large distances, where the exponential decay is dominant\cite{Cornelissen2015,Shan_2016}.

\begin{figure}[htbp]
\includegraphics[width=0.55\textwidth]{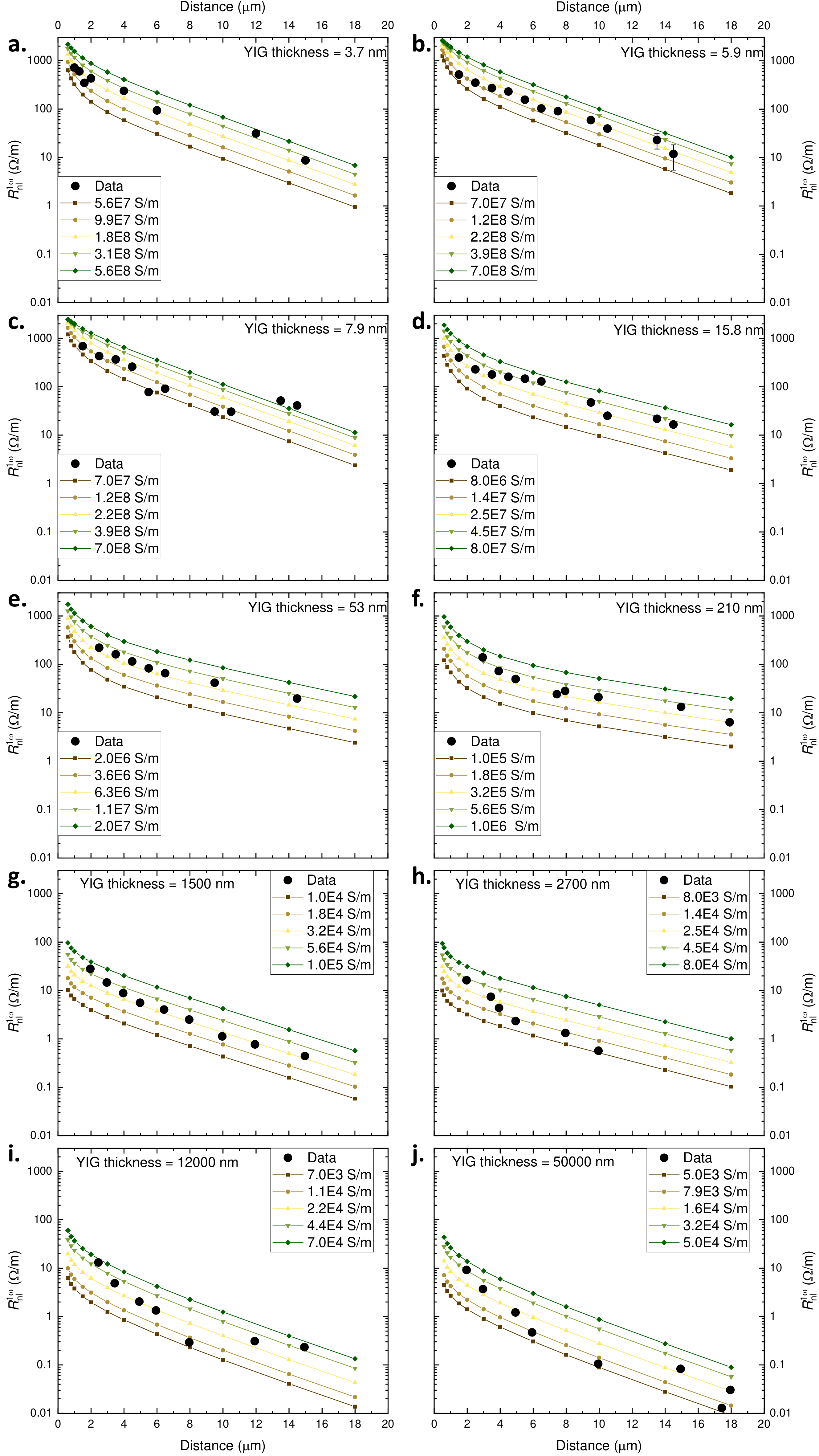}
\caption{Comparison between experimental data (black circles) and the
two-dimensional diffusive model obtained for different values of the magnon
spin conductivity $\sigma_{m}$, (lines brown to green). \textbf{a)} to
\textbf{j)} present the distance dependence of the non-local resistance for
different YIG thickness, from 3.7 nm up to 50 $\mu$m, indicated on the top of
each figure. Based on that comparison we obtain the value of the magnon spin
conductivity presented in Figure 5 of the main text.}%
\label{fig:simulations}%
\end{figure}
Each black circle in Figure \ref{fig:simulations} is an independent
measurement and susceptible to variations in the individual sample parameters.
It is clear from Figure \ref{fig:simulations} that the experiments can be well
fitted by a $\sigma_{m}$ that depends on $t_{\mathrm{YIG}}$.

\begin{table}[htbp]
\caption{Values of the magnon spin relaxation length adopted in the simulations.}%
\label{tab:spin_diffusion}%
\begin{ruledtabular}
			\begin{tabular}[c]{cc}
				YIG thickness [nm] & Spin relaxation length [$\mu$m]\\
				\colrule
				5.9 / 7.9 / 15.8 & 3.5\\
				53 & 6.0\\
				210 & 9.2\\
				1500 & 4.0\\
				2700 & 3.5\\
				12000 / 50000 & 4.0\\
			\end{tabular}
		\end{ruledtabular}
\end{table}

\section{Pt%
$\vert$%
YIG interface spin resistance}

\label{section:The role of the interface conductance}

\begin{figure}[htbp]
\includegraphics[width=0.55\textwidth]{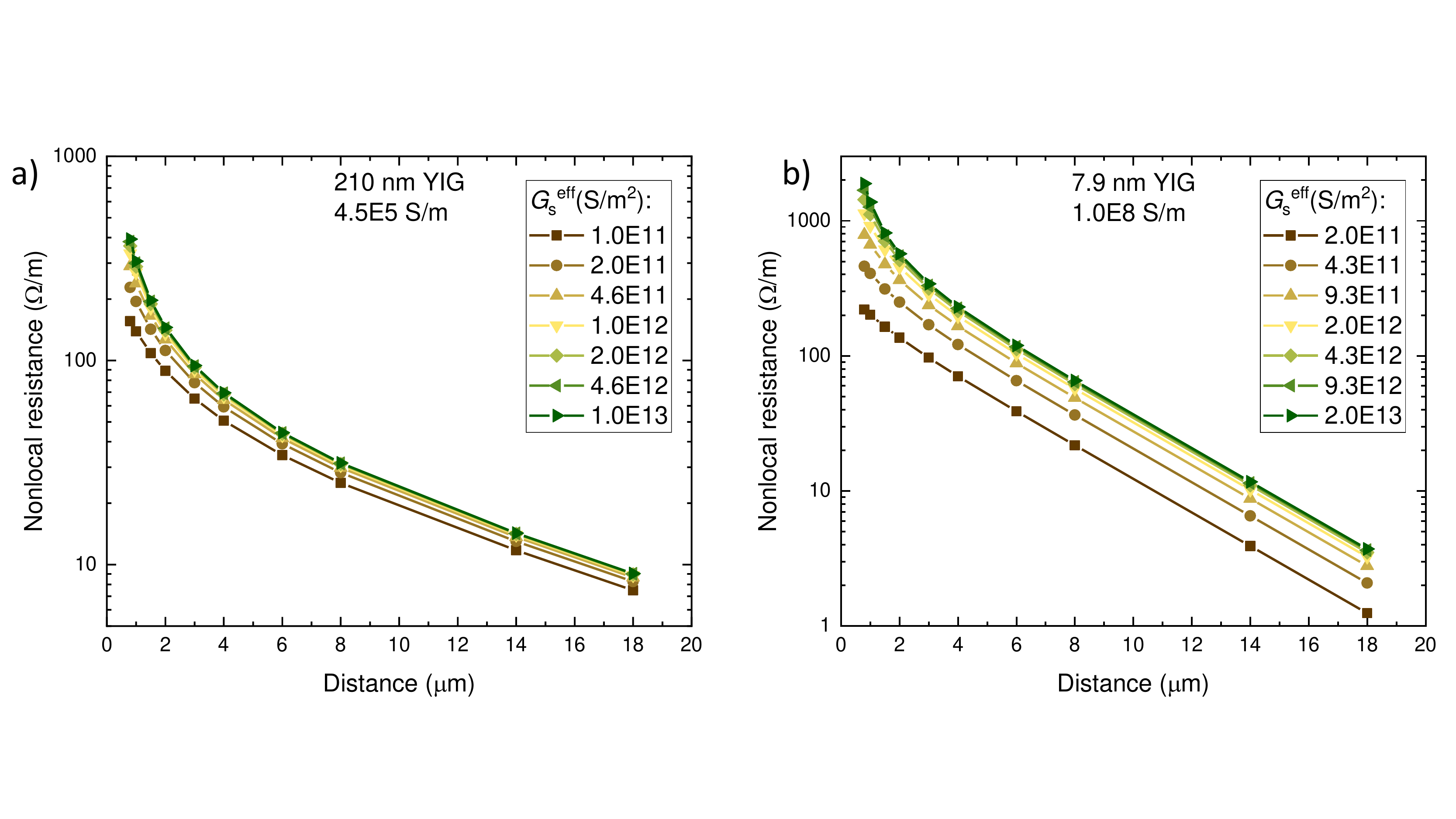}
\caption{Calculated non-local resistance for a range of values of the
effective spin conductance $G_{s}^{\mathrm{eff}}$ of a 400 nm wide injector/detector.
\textbf{a)} $t_{\mathrm{YIG}}=210\,$ nm, and \textbf{b)} $t_{\mathrm{YIG}%
}=7.9\,$nm YIG. In both cases, a significant change in the magnon transport
only occurs for $G_{s}^{\mathrm{eff}}<2\times10^{12}\,\text{S}/\text{m}^{2}$. }
\label{fig:interface_conduc}%
\end{figure}

\begin{figure}[htbp]
\vspace{0pt} \includegraphics[width=0.4\linewidth]{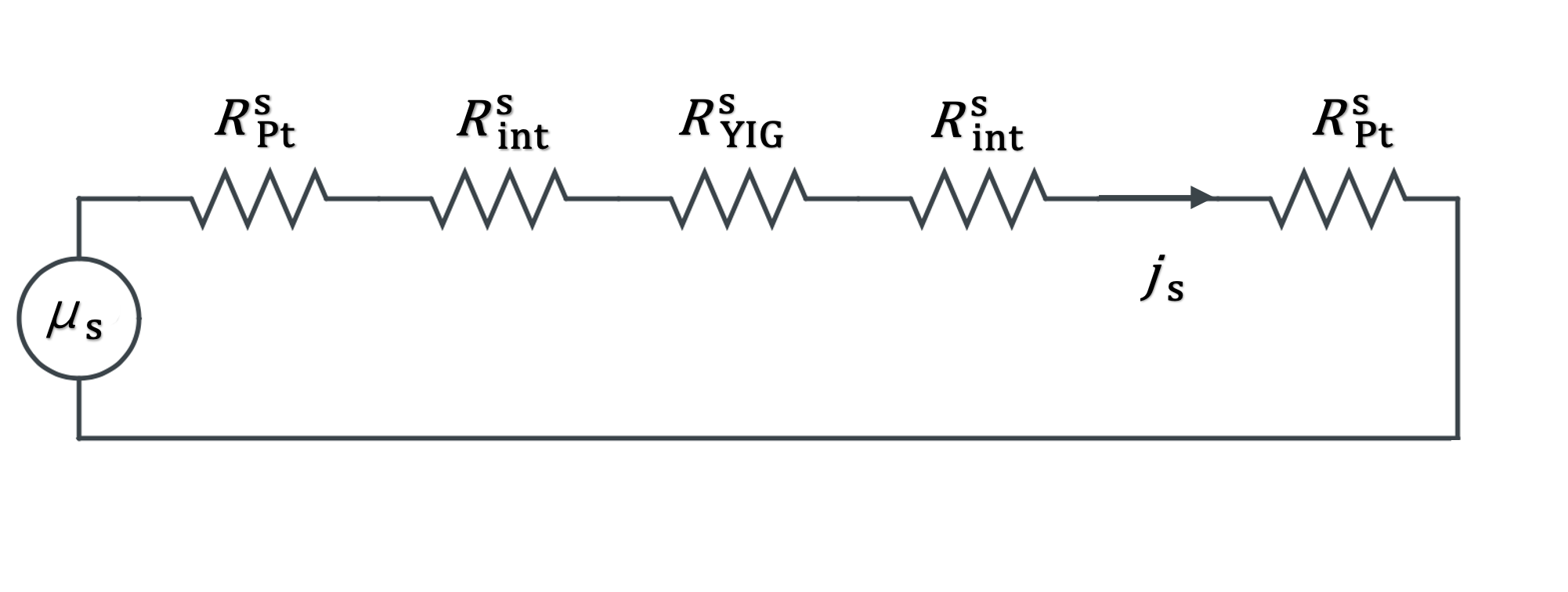}
\vspace{-10pt} 
\caption{This equivalent circuit model includes the relevant electronic and magnonic spin resistances and it is valid for short distance (within the magnon
relaxation length). $\mu_{s}$ is the spin accumulation induced by the
spin Hall effect. $R_{\mathrm{Pt}}^{s}$ is the spin resistance of the
Pt strip (injector/detector). $R_{\mathrm{int}}^{s}$ is the interface
spin resistance of Pt$\vert$YIG. $j_{s}$ is the electronic spin current
injected into the detector, and $R_{\mathrm{YIG}}^{s}$ is the spin
resistance of YIG, which is the parameter of interest obtained from the
two-dimensional model.}%
\label{fig:circuit_without_rel}%
\end{figure}

The low interface magnon resistance resulting from the large effective spin
mixing conductance makes the Pt strips act as a good spin source/sink.
Figure \ref{fig:interface_conduc} shows the calculated values of non-local
resistance for wide range of effective spin conductance, $G_{s}^{\mathrm{eff}}$. A
significant change in the non-local resistance occurs only when we suppress
$G_{s}^{\mathrm{eff}}$ to below $2\times10^{12}\ \text{S}/\text{m}^{2}$.
This value is lower than reported by Kohno et. al., $8.8\times 10^{12}\ \text{S}/\text{m}^{2}$\cite{surface_annealing} (after local annealing). In Figure \ref{fig:interface_conduc}, a higher $G_{s}^{\mathrm{eff}}$ does not significantly change the calculated non-local resistance, which represents that the giant increase of the non-local resistance we observed is not due to the increase of $G_{s}^{\mathrm{eff}}$.
Figure \ref{fig:circuit_without_rel} shows an equivalent circuit model omitting magnon relaxation. The spin current passing through the circuit is dominated by the spin conductance of YIG since $R_{\mathrm{int}}^{s}<R_{\mathrm{YIG}}^{s}$. Therefore, $R_{\mathrm{int}}^{s}$ can be disregarded and the
magnon transport measured by $R_{\textrm{nl}}$ is determined by the magnon spin
conductivity of YIG rather than the YIG%
$\vert$%
Pt interface conductance.


In an additional experiments we placed a third Pt strip between the injector and detector contacts. By absorbing magnons, the middle strip suppresses the non-local signal by a factor of 5 when $d=2\ \mathrm{\mu
}\text{m}$ and $t_{\mathrm{YIG}}=7.9\ \text{nm}$, and by a factor of 2 for
$d=4$ $\mu$m and $t_{\mathrm{YIG}}=53\ \text{nm}$. From this, we obtain a best fit with $G_{s}^{\mathrm{eff}}\approx2.5\times10^{12}\ \text{S}/\text{m}^{2}$ for
$t_{\mathrm{YIG}}=7.9\ \text{nm}$, and $G_{s}^{\mathrm{eff}}\approx1.0\times10^{12}%
\ \text{S}/\text{m}^{2}$ for $t_{\mathrm{YIG}}=53\ \text{nm}$. We therefore
adopted $G_{s}^{\mathrm{eff}}=2\times10^{12}\ \text{S}/\text{m}^{2}$ as the standard value
for the effective spin conductance in the other simulations.

\begin{figure}[htbp]
\vspace{0pt} \includegraphics[width=0.7\linewidth]{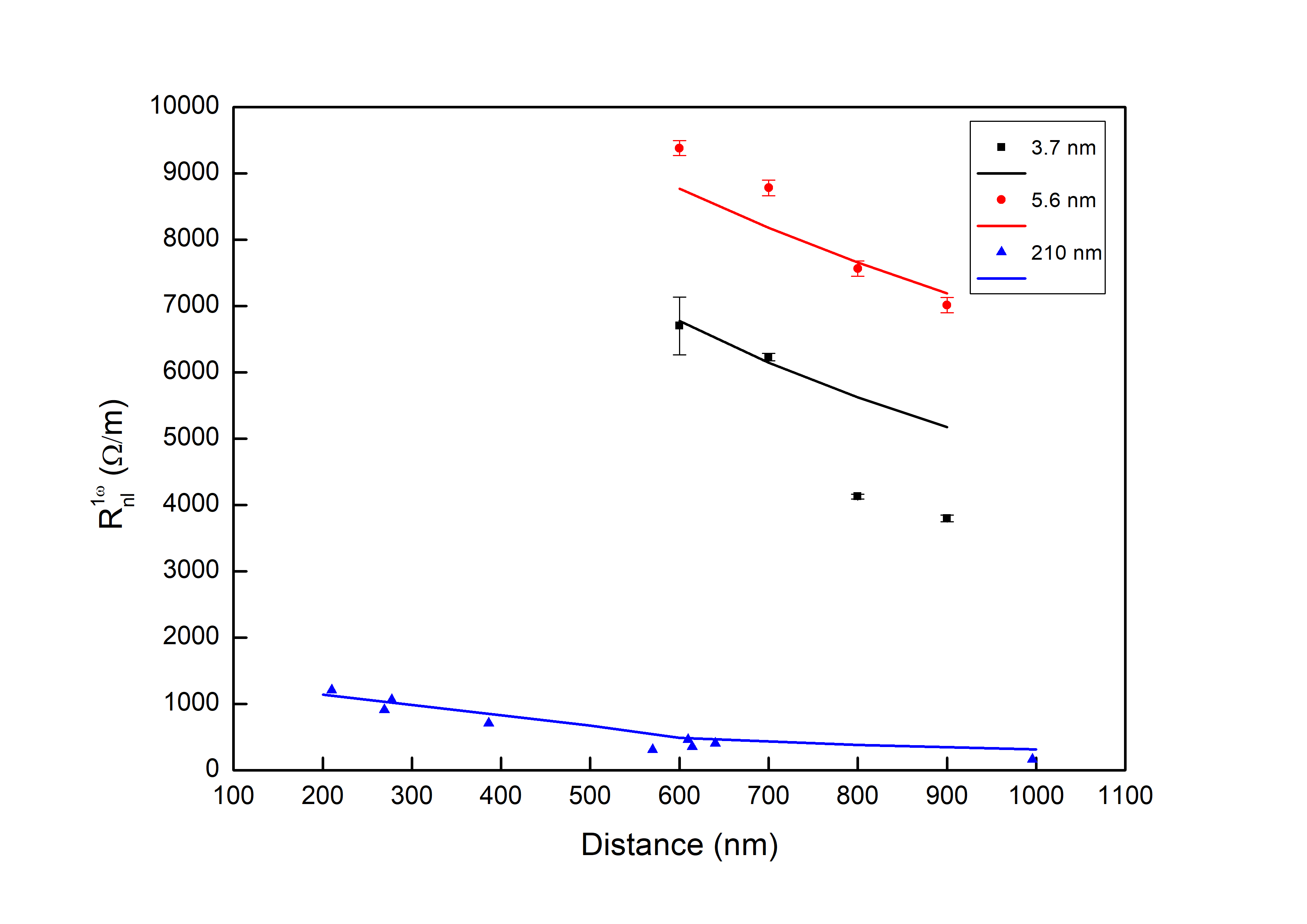}
\par
\vspace{-10pt}\caption{Non-local resistance as a function of injector-detector
separation distance for series B devices. The thickness of YIG films is 3.7 nm and 210 nm. The non-local resistance of 3.7 nm thickness YIG is more than
ten times larger than that in 210 nm thickness YIG at same center-to-center
distance. The lines are the simulation results based on a 2D-FEM model. The
magnon conductivities $\sigma_{m}$ here are 1.0$\times10^{8}$ S/m for 3.7
nm, 1.45$\times10^{8}$ S/m for 5.6 nm and 4.5$\times10^{5}$ S/m for 210 nm thickness
YIG.}%
\label{fig:e_together_100}%
\end{figure}

\begin{figure}[htbp]
\vspace{-10pt} 
\includegraphics[width=0.7\textwidth]{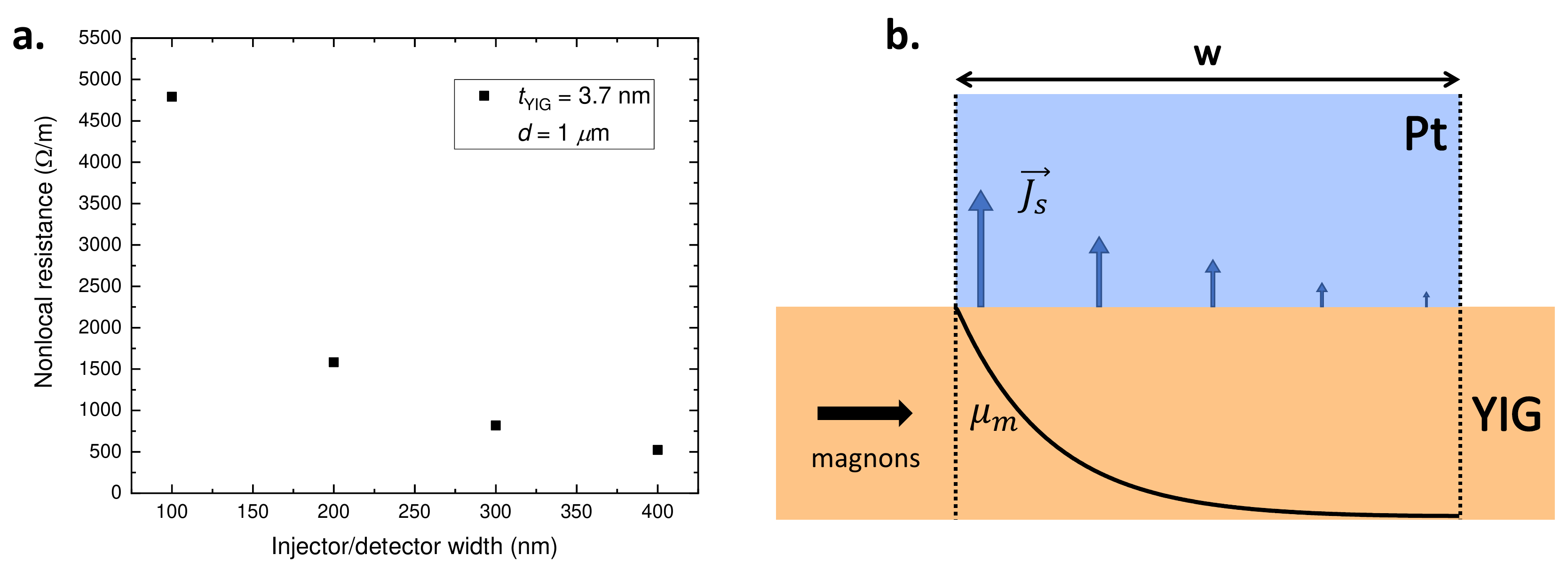}
\vspace{-10pt} 
\caption{\textbf{a)} Simulation for the non-local resistance with different
injector/detector width on 3.7 nm thickness YIG. The center-to-center distance
between the injector and the detector is 1 $\mu$m.
The simulations in Figure \ref{fig:width_differ_3_7} are in agreement with the experimental results in Figure \ref{fig:e_together_100}.
\textbf{b)} Schematic illustration of the decay of the magnon
chemical potential decay and spin current over the Pt width. }%
\label{fig:width_differ_3_7}
\end{figure}

We confirm the high $G_{s}^{\mathrm{eff}}$ by comparing results for different widths of the
injector/detector contacts. The non-local resistance from the Series B devices
in Figure \ref{fig:e_together_100} with width $w=100$\thinspace nm, is roughly
10 times higher than that of Series A devices with $w=400$\thinspace nm, which is mainly due to the higher values of the injector/detector conversion efficiency, Eq. \ref{eq:spinaccu}. The
simulations show that when $G_{s}^{\mathrm{eff}}$ is large, magnons are injected/absorbed
predominantly in/by only part of the interfaces, i.e., the magnon chemical
potential in the YIG film covered by the Pt strip decays according to an
absorption length $L_{\text{\textrm{absorb}}}=\sqrt{\sigma_{m}%
t_{\mathrm{YIG}}/G_{s}^{\mathrm{eff}}}$ as sketched in Figure
\ref{fig:width_differ_3_7} b) in terms of the interface chemical potential
$\mu_{m}$ and spin current, $\vec{J_{s}}$, in the Pt detector. Figure \ref{fig:width_differ_3_7} a) shows the
calculated resistance for $G_{s}^{\mathrm{eff}}=2\times10^{12}$ S/$\text{m}^{2}$, $\sigma
_{m}=1\times10^{8}$ S/m, $t_{\mathrm{YIG}}=3.7\,$nm and therefore
$L_{\mathrm{absorb}}=430$ nm that agrees with observations. This confirms our
value for $G_{s}^{\mathrm{eff}}$ even for the thinnest films.

\section{Magnetic properties of the YIG films}

\label{section:magnetic properties}

\begin{figure}[htbp]
\vspace{0pt} \includegraphics[width=0.8\linewidth]{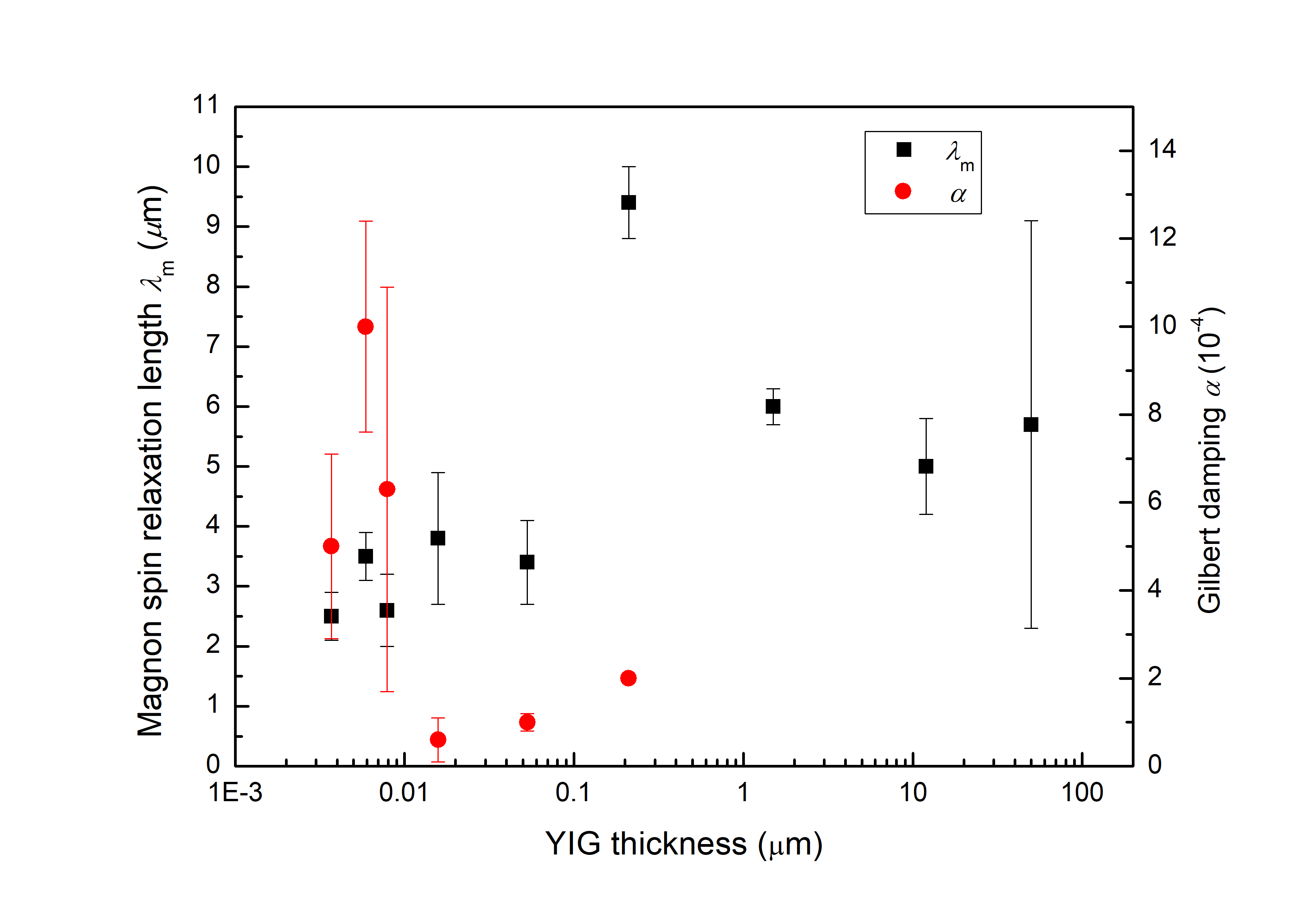}
\vspace{-10pt}\caption{Magnon spin relaxation length and Gilbert damping
parameter as a function of YIG film thickness. The thickness of YIG films
ranges from 3.7 nm to 50000 nm (shown on a logarithmic scale). The results
from 210 nm to 50000 nm thickness samples are adopted from Cornelissen et
al.\cite{Cornelissen2015} and Shan et al.\cite{Shan_2016}. The
Gilbert damping parameter of the YIG films is also presented here. Note that for
ultrathin YIG (around 10 nm), the magnon spin relaxation lengths are similar
to each other but the Gilbert damping parameters are quite different.}%
\label{fig:spin_length}%
\end{figure}

Table. \ref{gilbert} summarizes the magnetic properties of the YIG films are determined by broadband ferromagnetic resonance in the range of 2-40 GHz.  The magnon spin relaxation lengths obtained by fitting Eq. 3 in the main
text, are shown in Figure. \ref{fig:spin_length}. 

\begin{table}[htbp]
\caption{Magnetic properties of the YIG films by FMR\ characterization. }%
\label{gilbert}%
\vspace{3pt} \centering
\setlength{\extrarowheight}{-2pt}
\begin{tabular}
[c]{ccccc}%
\hhline{=====}
\vspace{0pt} &  &  &  &\\
YIG thickness  & Gilbert damping & Effective field & Inhomogenous linewidth &\\
$t_{\mathrm{YIG}}$ (nm)  & $\alpha$ ($10^{-4}$) & 
$H_{\text{eff}}=H_{\text{k}}-4\pi M_{\text{s}}$ (Oe) & $\Delta H_{\mathrm{inh}}$ (Oe) & \\\hline
\vspace{0pt} &  &  &  &\\
3.7 & $5\pm2.1$ & 1720 & 17&\\
\vspace{0pt} &  &  & &\\
5.6 & not available & 1900 & not available \\
\vspace{0pt} &  &  & &\\
5.9 & $10\pm2.4$ & 1950 & 43\\
\vspace{-0.5pt} &  &  & &\\
7.9 & $6.3\pm4.6$ & 1930 & 53\\
\vspace{-0.5pt} &  &  & &\\
15.8 & $0.6\pm0.5$ & 1960 & 7.9\\
\vspace{-0.5pt} &  &  & &\\
53 & $1.0\pm0.2$ & 1810 & 1\\
\vspace{0pt} &  &  & &\\
\hhline{====} &  &  & &
\end{tabular}
\end{table}

\section{The second harmonic non-local resistance}

\label{section:2w}

\begin{figure}[htbp]
\vspace{0pt} \includegraphics[width=1\linewidth]{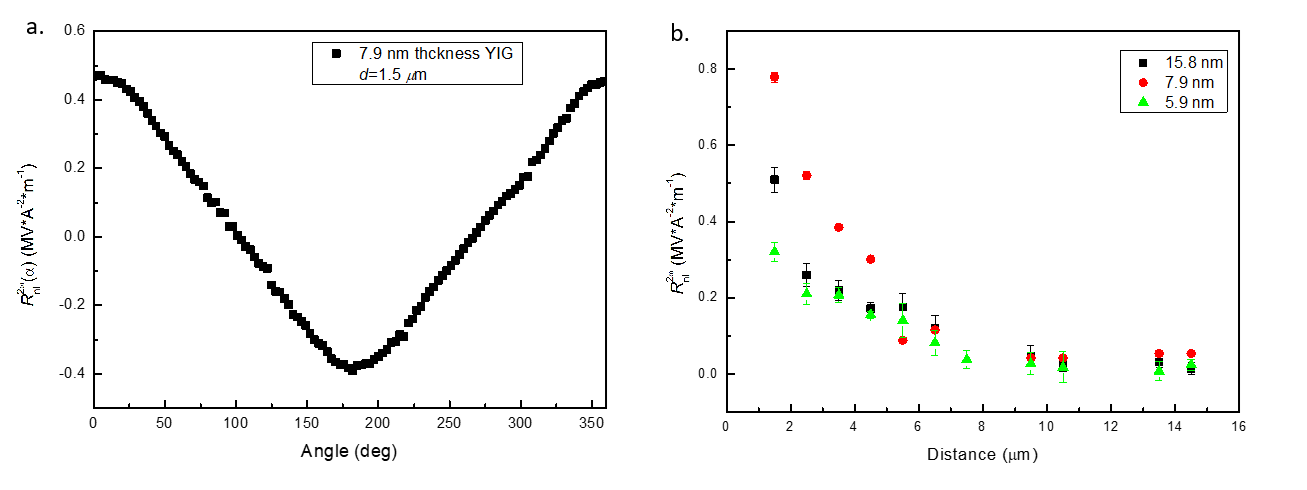}
\vspace{-10pt} \caption{The second harmonic non-local signals of ultrathin YIG
films. \textbf{a)} The angle-dependence for  $t_{\mathrm{YIG}}=7.9$\thinspace
nm and center-to-center distance of injector and detector $d=1.5\,\mathrm{\mu
m}$. \textbf{b)} Amplitude of the non-local signals as a function of $d$ for
the samples from series A.}%
\label{fig:2w}%
\end{figure}

The second harmonic signal in the non-local resistance is a measure of the
thermal spin generation in the magnetic film by the spin Seebeck effect.
Figures. \ref{fig:2w} a) shows results for this $R_{\mathrm{nl}}^{2\omega}$ ($\alpha$)
as a function of the direction of in-plane magnetic field. $R_{\mathrm{nl}}^{2\omega}$ can be extracted by
\begin{equation}
R_{\mathrm{nl}}^{2\omega}(\alpha)=1/2{R_{\mathrm{nl}}^{2\omega}}\cos
\alpha+{R_{0}^{2\omega}},\label{angle}%
\end{equation}
where $\alpha$ is the in-plane angle of $\mathbf{H}_{\mathrm{ex}}$ with the
$x$-axis. $R_{0}^{2\omega}$ is an offset signal, possibly by an unintended
conventional Seebeck voltage in the detector\cite{2w_offset}. We plot the
observed amplitude $R_{\mathrm{nl}}^{2\omega}\ $as a function of contact
separation in Figures. \ref{fig:2w} b) for $t_{\mathrm{YIG}}=5.9\,$ nm,
7.9 nm, 15.9 nm (series A with 400 nm wide injector/detector
strips).
$R_{\mathrm{nl}}^{2\omega}$ in ultrathin YIG is significantly reduced compared to that in thicker YIG films\cite{Cornelissen2015,Shan_2016,PhysRevB.96.184427},
because the generation of thermal magnons by SSE due to the vertical temperature gradient become less effective.
Figures. \ref{fig:2w} b) show neither a simple Ohmic or exponential decay as a function of $d$, which indicates that the generation and transport of magnons is  complex and requires detailed modelling of the temperature profile.
Just like the magnon conductivity, the spin Seebeck coefficient should also depend on $t_{\mathrm{YIG}}$.
A reliable extraction of both parameters as a function of thickness from the second harmonic signals appears impossible at this this time.

\section{Magnetic field dependence of the non-local signal in ultrathin films}

\label{section:field}

\begin{figure*}[htbp]
\vspace{0pt} \includegraphics[width=0.8\linewidth]{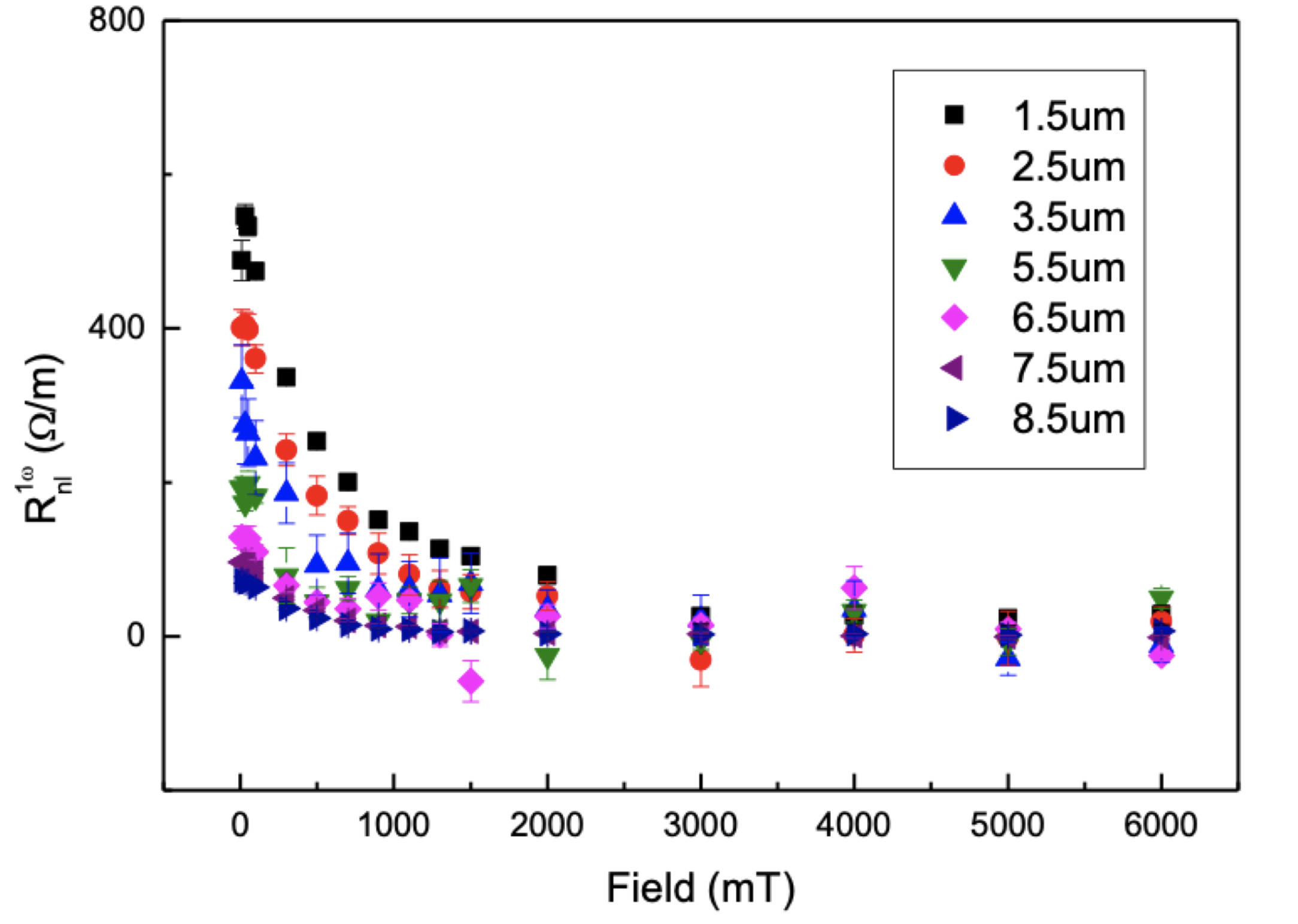}
\vspace{-10pt} \caption{The non-local resistance $R_{\text{\textrm{nl}}%
}^{1\omega}$ as a function of magnetic field and contact distance of the A
series sample with $t_{\mathrm{YIG}}=5.9\,$nm.}%
\label{fig:field5.9}%
\end{figure*}

\begin{figure}[htbp]
\vspace{0pt} \includegraphics[width=0.8\linewidth]{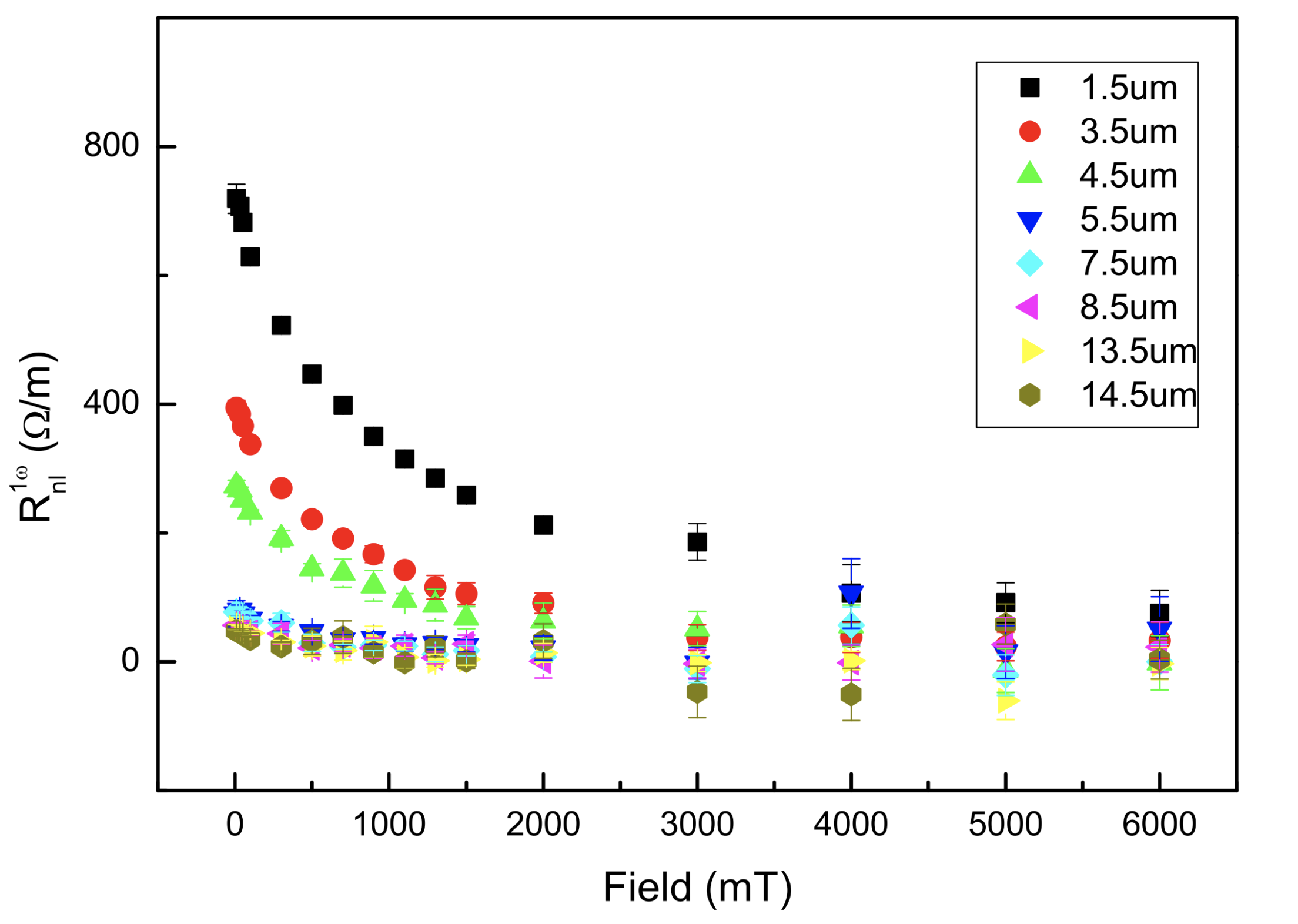}
\vspace{-10pt} \caption{The non-local resistance as function of magnetic field for $t_{\mathrm{YIG}%
}=7.9$\thinspace nm. }%
\label{fig:field7.9}%
\end{figure}

\begin{figure}[htbp]
\vspace{0pt} \includegraphics[width=1\linewidth]{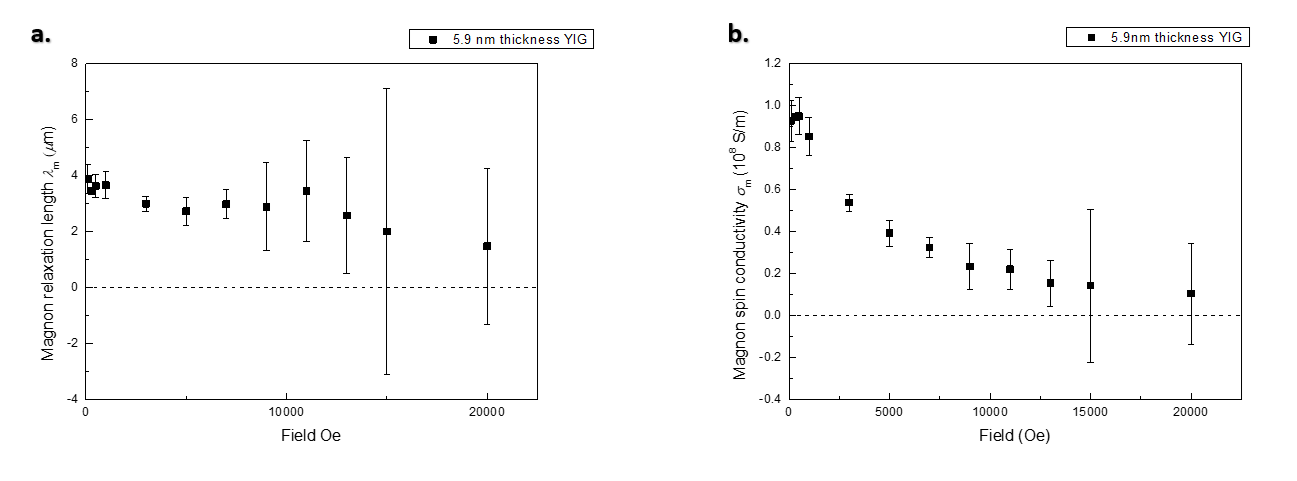}
\vspace{-10pt} \caption{$\lambda_{m}$ and $\sigma_{m}$ as a function of field
in a film with $t_{\mathrm{YIG}}=5.9\,$nm. \textbf{a)} The magnon relaxation
length $\lambda_{m}$  decreases slightly with increasing magnetic field.
\textbf{b)} The magnetic field strongly suppresses the magnon conductivity
$\sigma_{m}$..}%
\label{fig:field5.9f}%
\end{figure}

\begin{figure}[htbp]
\vspace{0pt} \includegraphics[width=1\linewidth]{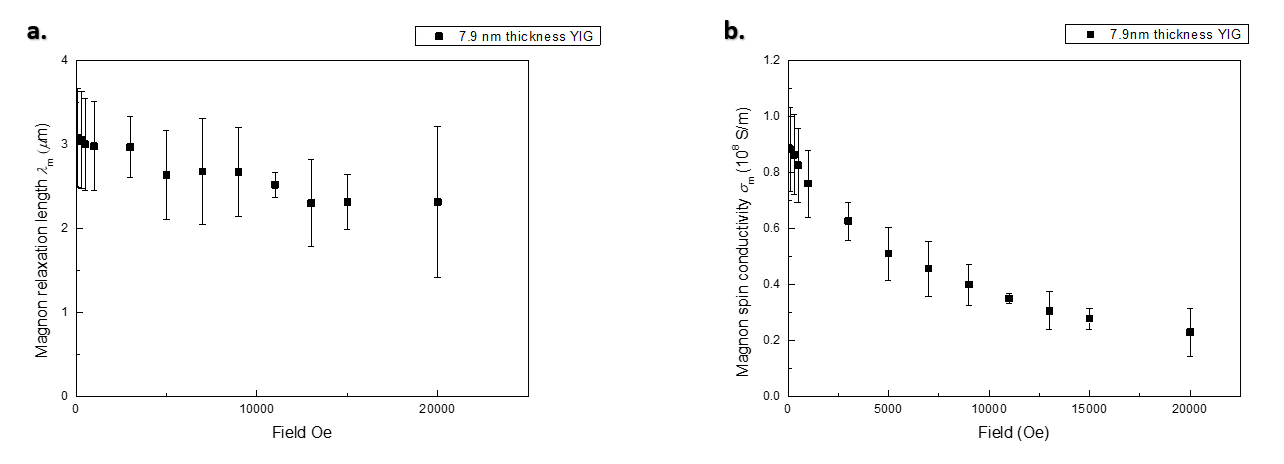}
\vspace{-10pt} \caption{\textbf{a)} $\lambda_{m}$ and \textbf{b)} $\sigma_{m}$, as a function of field
in a film with $t_{\mathrm{YIG}}=7.9$ nm.}%
\label{fig:field7.9f}%
\end{figure}

We measured the non-local transport in the series A devices with
$t_{\mathrm{YIG}}=5.9$\thinspace nm (Figure.$\,$\ref{fig:field5.9}) and
$t_{\mathrm{YIG}}=7.9$\thinspace nm (Figure.$\,$\ref{fig:field7.9}) as a
function of  the strength of a magnetic field along $x$ (in-plane, normal to
the contacts). We use Eq. 3 from the main text to extract the field-dependence
of the magnon conductivity and relaxation length, as shown in Figure.
\ref{fig:field5.9f} and Figure. \ref{fig:field7.9f}, where we use
$\eta_{\mathrm{cs}}=\eta_{\mathrm{sc}}=0.05$\thinspace $\Omega$ from Section
\ref{section:finite_element_modelling}.

While the magnon spin relaxation length only slightly decreases, the
increasing field suppresses the magnon spin conductivity stronger for the
thinner samples. This is in line with the dominance of the lowest magnon
subband in the thinnest samples, which is more susceptible to magnetic freeze-out of
the thermal magnon population.   

\newpage

\section{Temperature dependence of the non-local signal in ultrathin films}

\label{section:temperature} 

\begin{figure}[htbp]
\vspace{0pt} \includegraphics[width=0.7\linewidth]{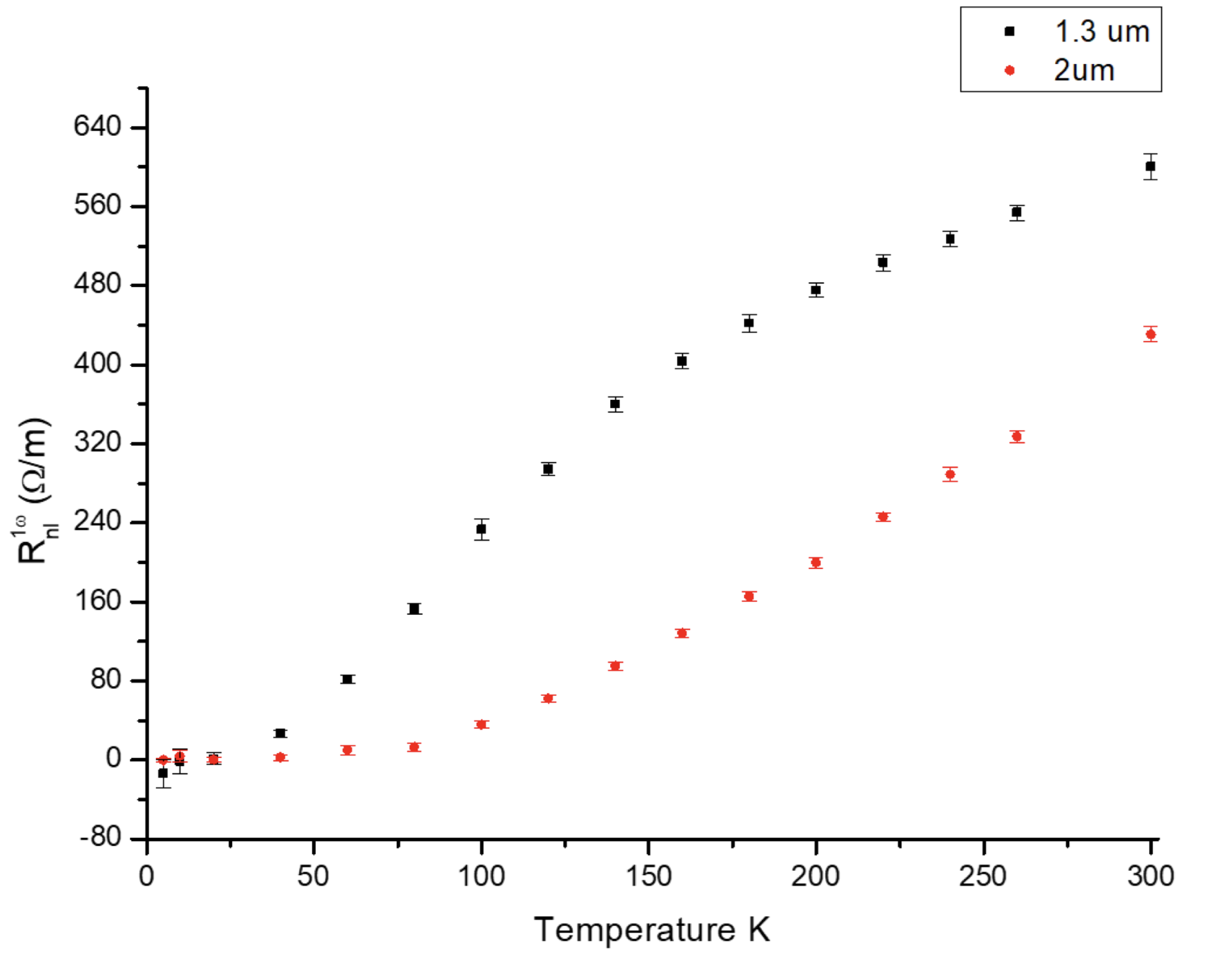}
\vspace{-10pt} \caption{The non-local resistance $R_{\mathrm{nl}}^{1\omega}$
as a function of temperature for different\ contact distances on a YIG film
with $t_{\mathrm{YIG}}=3.7$\thinspace nm. }%
\label{fig:temp3.7}%
\end{figure}

\begin{figure}[htbp]
\vspace{0pt} \includegraphics[width=0.8\linewidth]{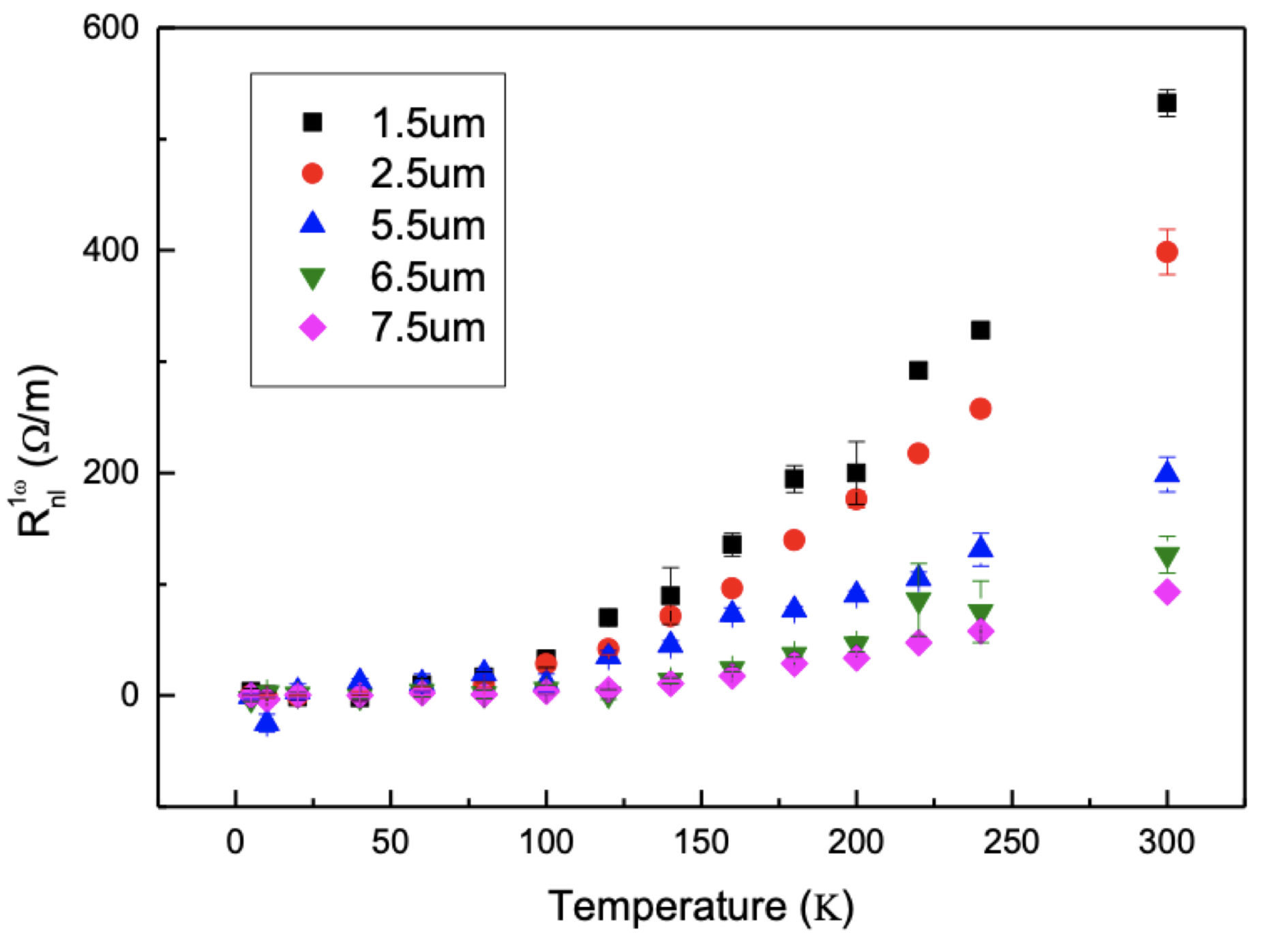}
\vspace{-10pt} \caption{Non-local resistance as a function of temperature for $t_{\mathrm{YIG}%
}=5.9$\thinspace nm.}%
\label{fig:temp5.9}%
\end{figure}

\begin{figure}[htbp]
\vspace{0pt} \includegraphics[width=0.8\linewidth]{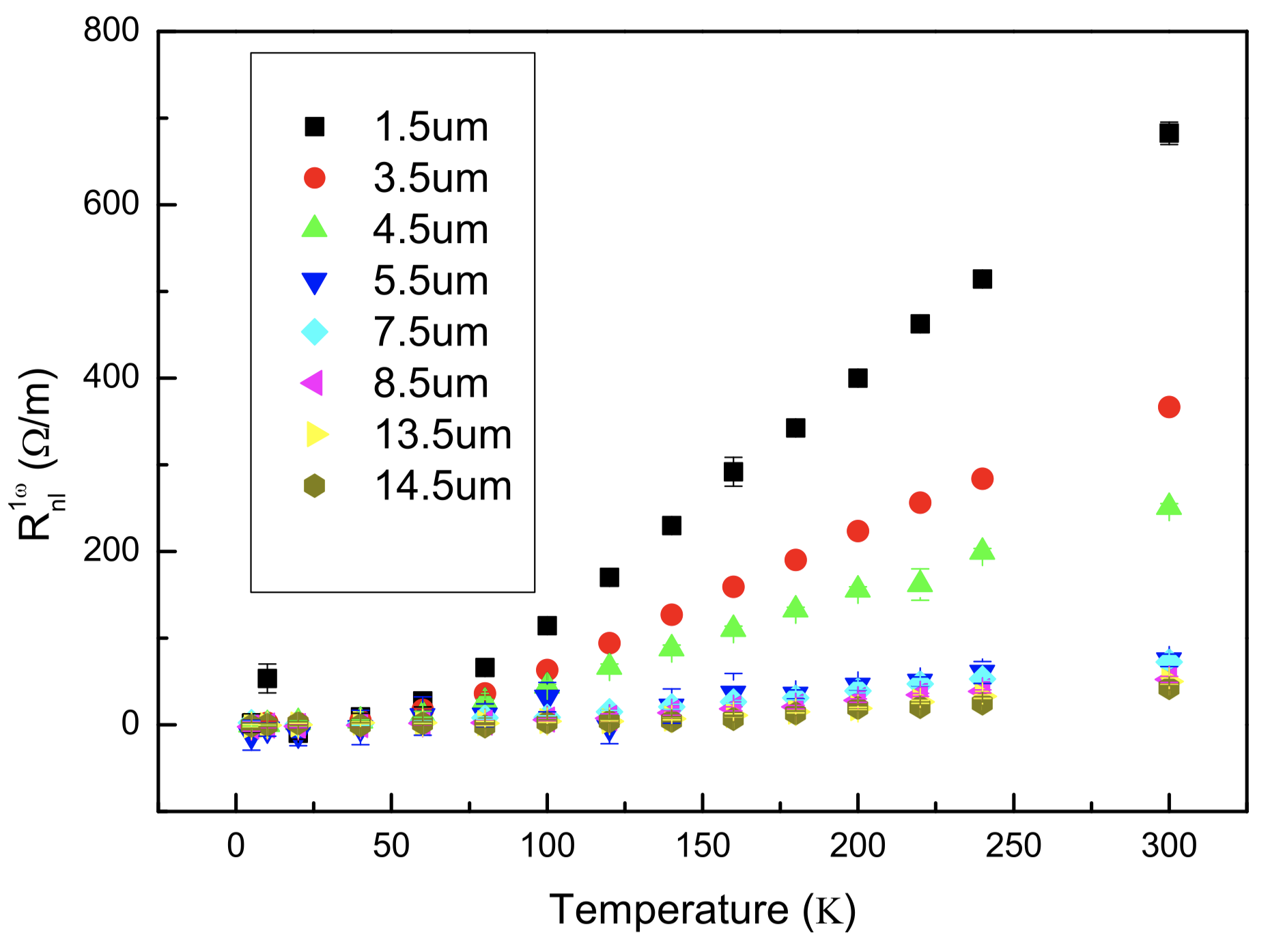}
\vspace{-10pt} \caption{Non-local resistance as a function of temperature for $t_{\mathrm{YIG}%
}=7.9$\thinspace nm.}%
\label{fig:temp7.9}%
\end{figure}

We measured the temperature dependence of the
non-local signal on series A samples with $t_{\mathrm{YIG}}=3.7$\thinspace nm
(Figure.$\,$\ref{fig:temp3.7}), $5.9$\thinspace nm (Figure.$\,$%
\ref{fig:temp5.9}) and $7.9$\thinspace nm (Figure.$\,$\ref{fig:temp7.9}) and a
magnetic field of 50 mT. $R_{\mathrm{nl}}^{1\omega}$ for $t_{\mathrm{YIG}%
}=3.7$\thinspace nm at low temperatures is more than five times larger for
$d=1.3\,\/\mathrm{\mu}$m than for $d=2\,\/\mathrm{\mu}$m. Since the decrease of $G_{s}^{\mathrm{eff}}$ due to the temperature is equivalent for both devices. The large change of the ratio of
$R_{\mathrm{nl}}^{1\omega}$ for $d=1.3\,\/\mathrm{\mu}$m and $d=2\,\/\mathrm{\mu}$m suggests that $R_{\mathrm{nl}}^{1\omega}$ is dominated by the magnon spin conductivity of YIG and not the
Pt$|$YIG interface spin conductance, which supports the conclusions from
Section \ref{section:The role of the interface conductance}.

\begin{figure}[htbp]
\vspace{0pt} \includegraphics[width=1\linewidth]{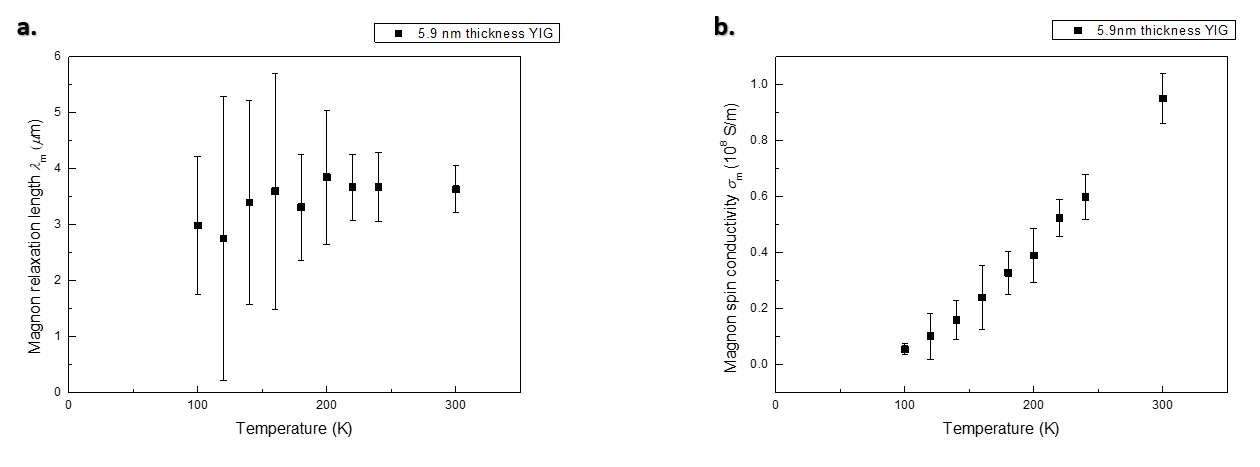}
\vspace{-10pt} \caption{\textbf{a)} $\lambda_{m}$ and \textbf{b)} $\sigma_{m}$ as a function of
temperature for $t_{\mathrm{YIG}}=5.9$\thinspace nm. }%
\label{fig:temp5.9f}%
\end{figure}

\begin{figure}[htbp]
\vspace{0pt} \includegraphics[width=1\linewidth]{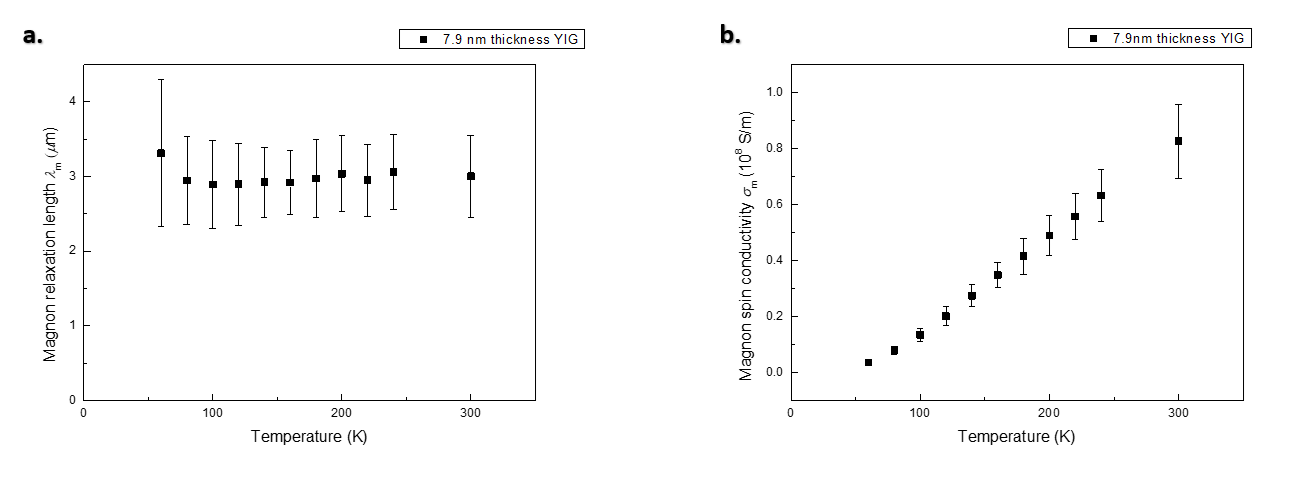}
\vspace{-10pt} \caption{\textbf{a)} $\lambda_{m}$ and \textbf{b)} $\sigma_{m}$ as a function of
temperature for $t_{\mathrm{YIG}}=7.9$\thinspace nm. }%
\label{fig:temp7.9f}%
\end{figure}

Results of the fit by Eq. 3 from the main text are shown in
Figure.$\,$\ref{fig:temp5.9f} and Figure.$\,$\ref{fig:temp7.9f}. The magnon
relaxation length only slightly changes with temperature, in contrast to
 that $\sigma_{m}$ decreases strongly with temperature. In thick YIG films, $R_{\mathrm{nl}%
}\left(  T\right)  $ decreases slowly at high temperatures and faster at low
temperatures \cite{temp, PhysRevB.101.184420}, but the decrease at low
temperatures is much more pronounced for ultrathin thickness YIG. Also this
result is consistent with the dominant role of the lowest magnon subband in thermal
transport in the thinnest films, since the two-dimensional magnon gas is more
susceptible to the freeze out of carriers at low temperatures.

\clearpage

\section{Highest occupied exchange mode of YIG at room temperature}

\label{section:pssw} 

\begin{figure}[htbp]
\vspace{0pt} \includegraphics[width=0.6\linewidth]{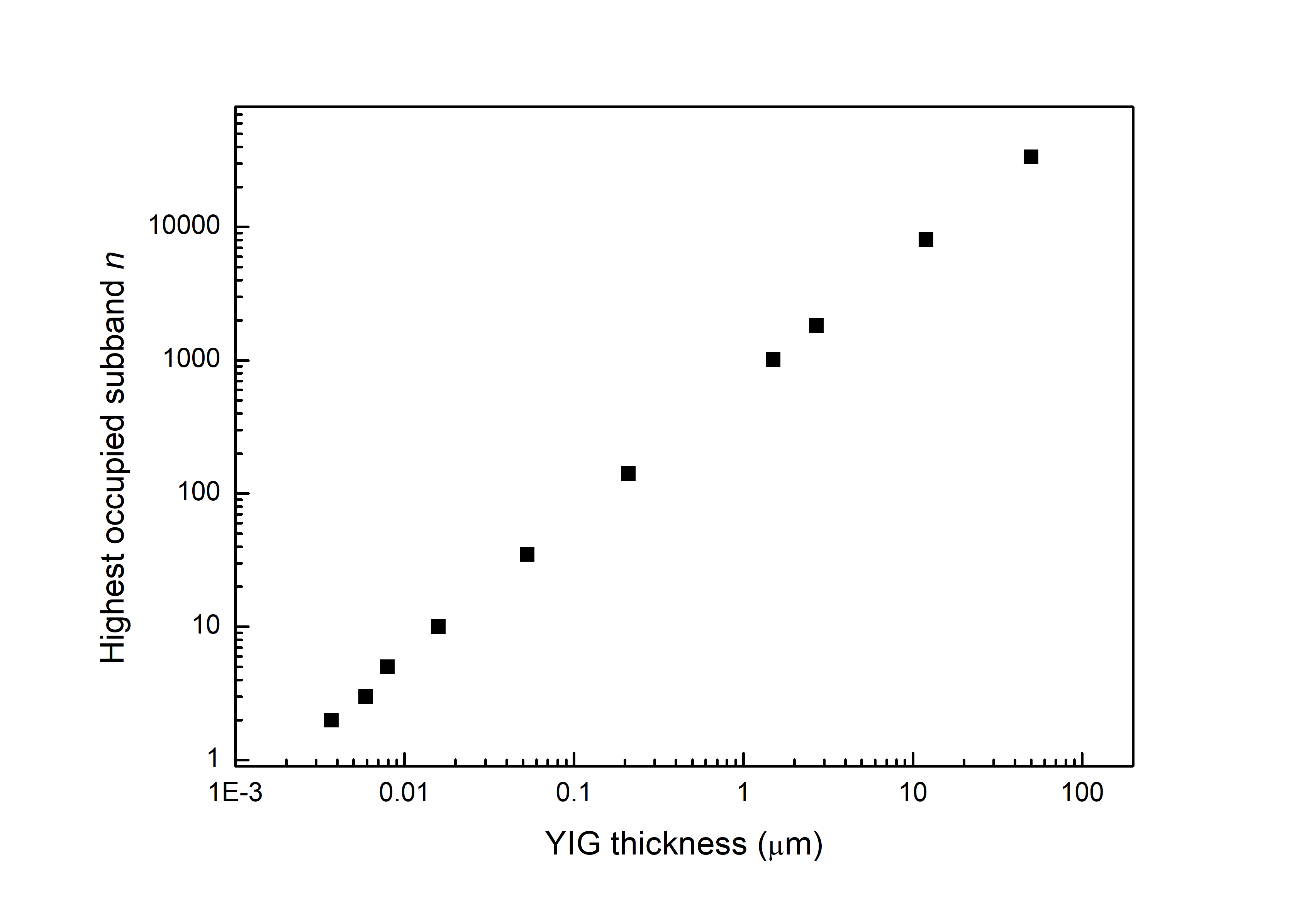}
\vspace{-10pt}\caption{Thickness dependency of the number of occupied
PSSW modes defined as the highest occupied subband $n$ at 300 K.}%
\label{fig:pssw}%
\end{figure}

Figure. \ref{fig:pssw} shows the highest occupied exchange perpendicular standing spin waves (PSSW) 2D subbands $n$ for different thickness YIG films at room temperature. The number of the occupied subbands is $n+1$. It is reduced from over $10^{4}$ to a few when the thickness of the films is down to 3.7 nm.

